\documentclass[twocolumn]{aastex61}

\usepackage{graphicx,amsmath,amsfonts,amssymb,subfigure,natbib,hyperref,rotating}
\newcommand{\fsps}{\texttt{FSPS}}

\newcommand{\mname}{\texttt{Prospector-$\alpha$}}

\newcommand{\prospector}{\texttt{Prospector}}

\newcommand{\angstrom}{\mbox{\normalfont\AA}}

\newcommand{\tuniv}{$t_{\mathrm{univ}}$}

\newcommand{\msun}{M$_{\odot}$}
\newcommand{\alphad}{$\alpha_{\mathrm{D}}$}

\begin{document}

\title{How to Measure Galaxy Star Formation Histories II: Nonparametric Models}

\correspondingauthor{Joel Leja}
\email{joel.leja@cfa.harvard.edu}

\author[0000-0001-6755-1315]{Joel Leja}
\affil{Harvard-Smithsonian Center for Astrophysics, 60 Garden St. Cambridge, MA 02138, USA}
\affil{NSF Astronomy and Astrophysics Postdoctoral Fellow}

\author[0000-0002-9280-7594]{Adam C. Carnall}
\affil{SUPA, Institute for Astronomy, University of Edinburgh, Royal Observatory, Edinburgh EH9 3HJ, UK}

\author[0000-0002-9280-7594]{Benjamin D. Johnson}
\affil{Harvard-Smithsonian Center for Astrophysics, 60 Garden St. Cambridge, MA 02138, USA}

\author[0000-0002-1590-8551]{Charlie Conroy}
\affil{Harvard-Smithsonian Center for Astrophysics, 60 Garden St. Cambridge, MA 02138, USA}

\author[0000-0003-2573-9832]{Joshua S. Speagle}
\affil{Harvard-Smithsonian Center for Astrophysics, 60 Garden St. Cambridge, MA 02138, USA}


\begin{abstract}
Nonparametric star formation histories (SFHs) have long promised to be the ``gold standard'' for galaxy spectral energy distribution (SED) modeling as they are flexible enough to describe the full diversity of SFH shapes, whereas parametric models rule out a significant fraction of these shapes {\it a priori}. However, this flexibility is not fully constrained even with high-quality observations, making it critical to choose a well-motivated prior. Here, we use the SED-fitting code \prospector{} to explore the effect of different nonparametric priors by fitting SFHs to mock UV-IR photometry generated from a diverse set of input SFHs. First, we confirm that nonparametric SFHs recover input SFHs with less bias and return more accurate errors than do parametric SFHs. We further find that, while nonparametric SFHs robustly recover the overall shape of the input SFH, the primary determinant of the size and shape of the posterior star formation rate (SFR) as a function of time is the choice of prior, rather than the photometric noise. As a practical demonstration, we fit the UV-IR photometry of $\sim$6000 galaxies from the GAMA survey and measure inter-prior scatters in mass (0.1 dex), SFR$_{100\; \mathrm{Myr}}$ (0.8 dex), and mass-weighted ages (0.2 dex), with the bluest star-forming galaxies showing the most sensitivity. An important distinguishing characteristic for nonparametric models is the characteristic timescale for changes in SFR(t). This difference controls whether galaxies are assembled in bursts or in steady-state star formation, corresponding respectively to (feedback-dominated/accretion-dominated) models of galaxy formation and to (larger/smaller) confidence intervals derived from SED-fitting. High-quality spectroscopy has the potential to further distinguish between these proposed models of SFR(t).
\end{abstract}
\keywords{
galaxies: fundamental parameters
}
\section{Introduction}
Observational constraints on galaxy star formation histories (SFHs) enable us to understand many aspects of galaxy formation. The SFHs of quiescent galaxies provide insight into their formation conditions in the early universe and also into the mechanism by which star formation is quenched \citep{thomas05,graves08,conroy14,choi14,pacifici16,carnall18,schreiber18}. The SFHs of star-forming galaxies inform us about the timescales on which star formation rates (SFRs) change and the rate at which stellar mass is assembled on the star-forming sequence \citep{gallazzi05,panter07,leitner12}. Measuring the histories of low-mass dwarf galaxies provides insights into physics and timing of reionization \citep{tolstoy09,weisz14b,weisz14c}. Finally, modeling of galaxy SFHs is necessary to accurately measure stellar masses, star formation rates (SFRs), metallicities, and dust contents from their observed spectral energy distributions (SEDs) \citep{bell01,wuyts11a,conroy13a,leja17}.

Though SFHs in realistic galaxies can be arbitrarily complex, they are often modeled with simple functional forms. These forms are computationally fast and conceptually straightforward. The most common is an exponential form with $\mathrm{SFR} \propto t^{\alpha}e^{-t/\tau}$, with $\alpha=0$ known as an exponentially declining SFH and $\alpha=1$ known as a delayed-exponentially declining SFH. As our picture of galaxy formation becomes more complex, additional forms have been adopted to better describe the range of behaviors for SFR(t). These forms include rising SFHs \citep{buat08,maraston10,papovich11}, lognormals \citep{gladders13,abramson15,diemer17}, double power-laws \citep{carnall18}, and exponentially declining SFHs modified to decouple late-time and early-time SFRs \citep{simha14,ciesla16}.
 
These complex new functional forms can describe the majority of simulated galaxy SFHs quite well \citep{simha14,diemer17}, implying that simulated SFHs are, on average, relatively smooth. However, even if a functional form can provide an excellent description of the SFH when fit directly to SFR(t), it can fail to recover the true SFH when fit to galaxy observations. This happens because typical observations will only weakly constrain the SFH; in this scenario, not only must the functional form be able to describe the true SFH, it must also downweight the large range of qualitatively different SFHs which can show similar levels of agreement with the data. This is challenging for complex parametric models as parametric formulas are often chosen to maximize the range of SFHs that can be described rather than being chosen for having well-behaved priors. Furthermore, comparisons to simulations also show that a significant minority of simulated galaxy SFHs have more complex behavior than can be described with parametric models. This behavior includes bursts of star formation, sudden quenching, and rejuvenation events. The inability to model this behavior can lead to catastrophic failures in recovered star formation histories and result in biases in derived masses and star formation rates \citep{simha14}.

A well-known solution to these problems is so-called ``nonparametric'' SFHs, defined as models which explicitly do not assume a functional form for SFR(t)\footnote{In some sense ``nonparametric'' is a misnomer as there are indeed parameters in a nonparametric model.}. The simplest nonparametric models fit directly for the mass formed in a series of piecewise constant functions (i.e., a step function) in time \citep{fernandes05,ocvirk06,kelson14b,leja17,chauke18}. Other nonparametric models use adaptive time binning (e.g., \citealt{tojeiro07}) or directly fit libraries of SFHs measured from theoretical models of galaxy formation \citep{finlator07,pacifici12}. Nonparametric models tend to be more computationally expensive than parametric models as they must generate more SEDs to constrain a wider range of behavior in SFR(t); in return, nonparametric models promise the capability to describe the full breadth of complexity in galaxy star formation histories.

These models have typically been reserved for very high-quality data as galaxy observations are only typically only weakly informative about stellar age (e.g., \citealt{conroy13a}). The seminal work of \citet{ocvirk06} demonstrates that the problem of SFH recovery is ill-conditioned, i.e. that a small error in the observations typically results in much larger errors in the recovered star formation histories. They solve this problem by regularizing the inversion from observations to SFHs, specifically by introducing an additional assumption that the solution is a smooth function of time. Using this formalism, \citet{ocvirk06} show that only about eight characteristic episodes of star formation history can be recovered with high-quality optical spectra of resolution $R = 10,000$, signal-to-noise ratio (S/N) = 100, and wavelength coverage $4000-6800$ \angstrom{}. 

In this work, we further explore the ability of nonparametric SFHs to extract information from observations, with a companion work \citep{carnall18b} performing an analogous exploration of parametric models. A key difference from \citet{ocvirk06} is that this work is performed in a Bayesian framework. The Bayesian prior probability distribution takes the place of regularization in confining the model to a series of predetermined `acceptable' solutions. Here we explore the effect of the adopted prior on the posterior probability distribution (hereafter `posterior') which is being sampled. We test a few different priors as there are many unique ways in which a nonparametric model can be constructed. The tests cover a range of possible priors, from a simple uniform prior over log(mass) formed in each bin to a continuity prior emphasizing smooth star formation histories. Unlike parametric formulae, it is easy to tune the prior for nonparametric SFHs tune to follow expectations from theoretical models of galaxy formation. In this way, nonparametric SFHs can be useful even with low-quality data such as noisy photometry, as it is with poor data that having a well-tuned prior is most important.

The paper is structured as follows. Section \ref{sec:priors} contains an introduction to the different nonparametric priors that are tested in this work. Section \ref{sec:mocks} describes how the mock galaxies are generated and fit. Section \ref{sec:results} presents results from fitting mock galaxy data with different nonparametric priors. Section \ref{sec:gama} describes the results of fitting UV-IR photometry from the Galaxy and Mass Assembly (GAMA) survey with two different nonparametric priors. Section \ref{sec:par_vs_nonpar} contrasts the performance of parametric and nonparametric SFH fits, while Section \ref{sec:sim_sfhs} discusses the link between simulations of galaxy formation and nonparametric SFH priors. Section \ref{sec:conclusion} concludes the paper.

Where applicable, we use a WMAP9 cosmology \citep{hinshaw13} and a Chabrier initial mass function \citep{chabrier03}. We always report the median of the posterior, and 1$\sigma$ error bars are the $16^{\mathrm{th}}$ and $84^{\mathrm{th}}$ percentiles.

\begin{figure*}[t!h!]
\begin{center}
\includegraphics[width=0.95\linewidth]{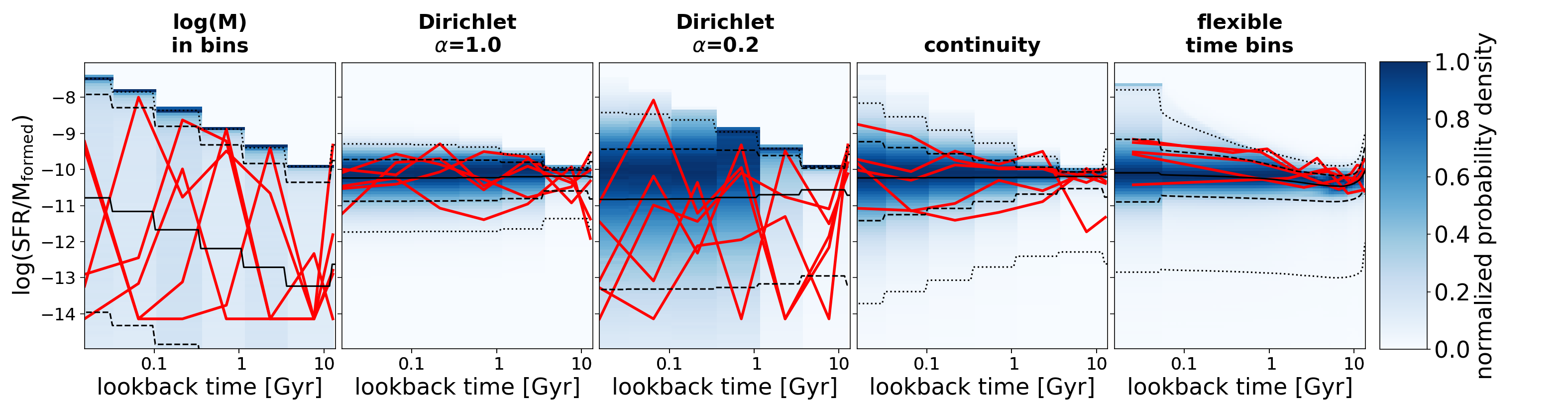}
\caption{Different choices for a nonparametric SFH prior produce different behavior in SFR(t). Each panel shows the prior probability distribution function (shaded in blue) for a different nonparametric SFH prior. The black (solid, dashed, dotted) lines mark the (median, 1$\sigma$, 2$\sigma$) levels of the distribution. Several individual draws are shown in red to illustrate the behavior of different priors. While all of these nonparametric priors are flexible enough to describe an arbitrary shape in SFR(t), they emphasize different behaviors. For example, the log(M) and Dirichlet $\alpha=0.2$ priors are weighted so as to produce multiple sharp quenching and rejuvenation events, while other priors select for star formation histories that are smooth in time.}
\label{fig:sfh_prior}
\end{center}
\end{figure*}

\section{Physical Model and Priors}
\label{sec:priors}
Here we describe the different priors explored in this work. As a brief introduction to Bayesian terminology, Bayes theorem states:
\begin{equation}
P(M|D) = \frac{P(D|M)P(M)}{P(D)}
\end{equation}
where $D$ is the data and $M$ is the physical model. In the Bayesian interpretation, $P(M)$ (the prior) is the estimate of the probability of the model $before$ comparison with the data. $P(M|D)$ (the posterior) is the probability of the model $after$ comparison with the data. $P(D|M)$ (the likelihood) is the probability of the measuring data given the model. Finally, $P(D)$ (the model evidence) is a normalizing factor used for comparison between different models.
\subsection{A Piecewise Constant Model for SFR(t)}
\label{sec:the_model}
In this work, we explore the effect of different nonparametric SFH priors P(M) on the resulting physical posteriors. In practice, each nonparametric prior under investigation is allotted $N=7$ parameters. These parameters simultaneously specify 7 distinct time-resolution elements and the overall normalization of the SED. Most of the nonparametric priors in this work additionally require specifying fixed time bins as input. For consistency, we adopt the following time bins in all such models:
\begin{align}
\begin{split}
\label{eqn:bins}
0 <\:  &t < 30\:  \mathrm{Myr} \\
30 <\:  &t < 100\:  \mathrm{Myr} \\ 
100 <\:  &t < 330\:  \mathrm{Myr} \\ 
330 \: \mathrm{Myr} < \: &t < 1.1\:  \mathrm{Gyr} \\
1.1 <\:  &t < 3.6\:  \mathrm{Gyr} \\ 
3.6 <\:  &t < 11.7\:  \mathrm{Gyr} \\ 
11.7 <\:  &t < 13.7\:  \mathrm{Gyr}
\end{split}
\end{align}

Aside from the first two bins and the last bin, these are spaced equally in logarithmic time, following the finding of \citet{ocvirk06} that the distinguishability of simple stellar populations is roughly proportional to their separation in logarithmic time. The last bin is deliberately much smaller to permit a maximally old population in all models. 

In Appendix \ref{appendix:tbins}, we explore varying the number of time bins between $N_{\mathrm{bins}} =4-14$ and show that the results of the mock analysis are largely insensitive to the number of bins as long as N$_{\mathrm{bins}} \gtrsim 4$. This highlights the fact that the Bayesian framework is robust to the classic dangers of fitting `overly complex' models, defined loosely as models which allow too much parameter space. In contrast to classical approaches where overly complex models result in `overfitting', i.e. overly tight constraints on parameters of interest, the outcome of allowing more parameter space than can be constrained by the data in a Bayesian framework is a {\it lack} of useful constraints on parameters of interest. Indeed this is often the desired outcome: if the model is not constrainable by the data, then the posteriors {\it should} return a null answer. Conversely, these tests also show that using $N\lesssim4$ time bins are typically insufficient to convey all of the necessary information in the data.

We note that there do exist methods to determine the appropriate number of bins on-the-fly, such as adaptively binning in time \citep{tojeiro07} or using evidence comparison to determine the optimal number of bins \citep{dye08,iyer19}. However, we argue that fitting many piecewise constant functions is more scalable in sampling framework for two complementary reasons. First, it is computationally much less expensive to run a fit with N$>$10 bins (Appendix \ref{appendix:tbins}) than it is to run two fits with any number of bins. Second, instead of using an always somewhat arbitrary statistical penalization to adjusting the model on the fly to match the data, it is more straightforward to include ``more bins than the data warrant'' and let the sampler fully map the inter-bin covariances allowed by the prior and the data. The potential failure mode for this is underfitting, which is caused by an excess of model flexibility and results in overestimated uncertainties. This danger can be mitigated by choosing a prior that weights for physically plausible forms of SFR(t) and weights against implausible forms, a complex problem which we spend the rest of the paper exploring.
\subsection{Priors}
\begin{figure*}[t!h!]
\begin{center}
\includegraphics[width=0.95\linewidth]{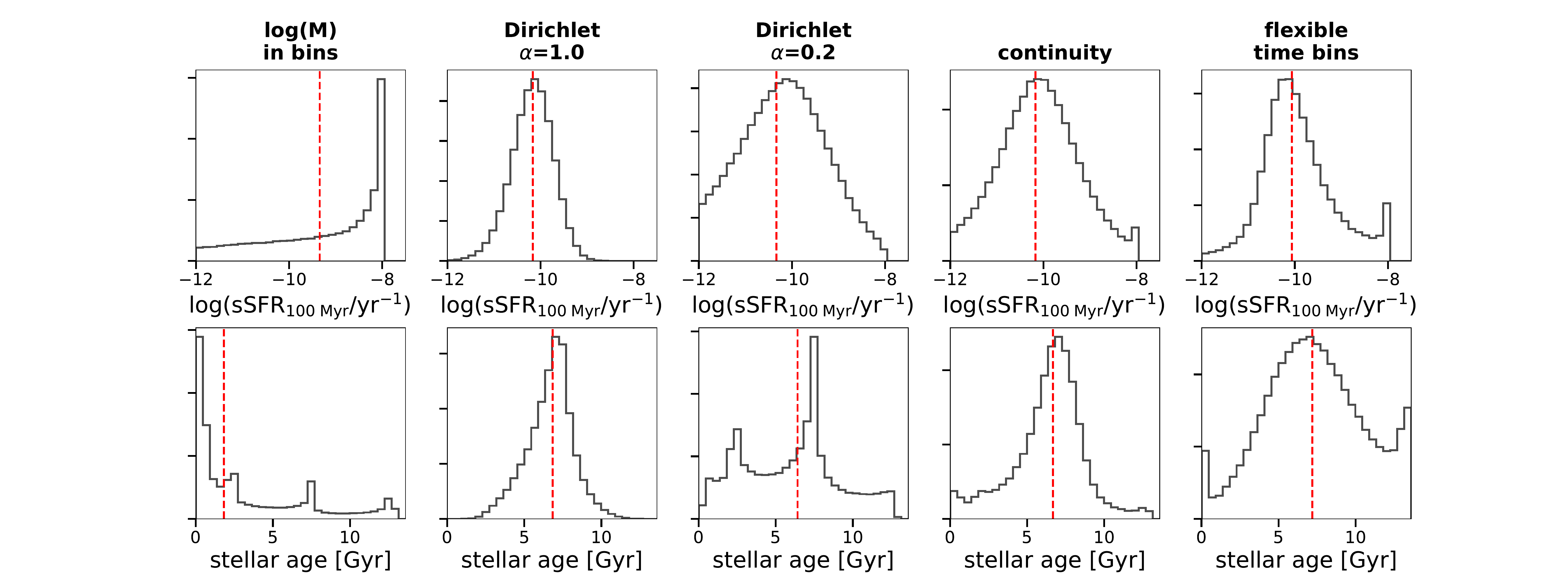}
\caption{The choice of SFH prior determines the prior on stellar age and specific star formation rate. The top row shows the prior on sSFR averaged over the most recent 100 Myr while the bottom row shows the mass-weighted age priors. The columns show different SFH priors. The red dashed line marks the median of the prior. These differences demonstrates that the choice of prior affect the derived ages and sSFRs. For example, priors which favor ``bursty" SFHs (logM, Dirichlet $\alpha=0.2$) show clear discretization effects related to the choice of time bins. Furthermore, the logM prior strongly prefers rising star formation histories. Averaging over the most recent 100 Myr causes the truncation in the sSFR histogram at log(sSFR) = -8.}
\label{fig:par_prior}
\end{center}
\end{figure*}

The prior probability function for each nonparametric prior is shown in Figure \ref{fig:sfh_prior}, and these priors projected into mass-weighted age and specific star formation rate space are shown in Figure \ref{fig:par_prior}.

\subsubsection{LogM prior}
\label{sec:logm_prior}
The most straightforward and popular nonparametric model fits for the mass formed in $N$ fixed time bins \citep{walcher15,belli18,morishita18}. A version of this method has also been used in STARLIGHT \citep{fernandes05} and in analysis of the CSI \citep{kelson14b} and LEGA-C surveys \citep{chauke18}, though these works report best-fit rather than marginalized SFHs and fit for linear rather than logarithmic weights.

The prior on SFR(t) imposed by the logM prior is shown in Figure \ref{fig:sfh_prior}. The spacing of the time bins in equal logarithmic bins means the logM prior is not agnostic about the shape of SFR(t) but instead prefers $rising$ star formation histories and high instantaneous sSFRs. Furthermore, the logM prior tends to form the majority of the mass in one or two time bins: this can be seen clearly in the top-heavy conditional probability function for the SFR in any given time bin. This manifests itself in the mass-weighted age prior (Figure \ref{fig:par_prior}) as an overall preference for young ages and specific preference for ages corresponding to the center of the fixed time bins.

This formalism explicitly couples the normalization of the observed SED to the SFH parameterization. In practice, this can introduce difficulties in choosing the prior range for the mass formed in each time bin: using a constant lower limit for the mass formed will effectively put different ``floors'' on SFR(t) depending on the mass of the galaxy in question. For this work, we allow each mass formed in each bin to vary between $3 <$ log(M/\msun{}) $<12$, which is comfortably outside of the input mass of $\log(\mathrm{M/M}_{\odot})=10$ used in the mock tests.

Overall, the physical parameters resulting from a logM prior are sensitive to the choice of time bins and show clear preferences for rising star formation histories that are `bursty', showing multiple sharp quenching and rejuvenation events. This burstiness is a natural consequence of the geometry of the logM prior. This can be understood intuitively by imagining a series of random draws from the logM prior which add up to a fixed total mass by construction (in a typical fit, this total mass is loosely constrained by the normalization of the SED). Most of the SFHs thus constructed will have significant mass in only one or two time bins: this naturally results in `bursty' SFHs. The logM prior does have a clear advantage over the other models: it has the widest probability distribution and thus tends to be the most flexible. However, it is shown in Section \ref{sec:results} that this flexibility is not well constrained by typical galaxy observations, even at very high S/N.

We have also explored several models closely related to the logM prior which we describe briefly here with no further analysis. Replacing the logM model with a logSFR model (i.e. fitting for the SFR instead of the mass) successfully removes the bias towards rising SFHs while retaining the other characteristics of the logM prior. Fitting instead for the $linear$ instead of logarithmic mass in fixed time bins exacerbates the downsides of the logM prior -- specifically a strong preference for young ages and forming the majority of the mass in one or two bins -- while showing no clear upside. This is consistent with the argument in \citet{simpson17} that flat priors on a parameter spanning several orders of magnitude are actually strongly informative, and, in this case, the logarithmic prior is the minimally informative choice. The linear prior further introduces significant convergence issues for the sampler (sampler described in Section \ref{sec:mocks}) as there are very large regions of parameter space which have essentially flat probability: for example, for a $10^{11}$ \msun{} galaxy, there is no practical difference in the photometry between forming 10$^5$ and $10^9$ solar masses in the oldest time bin. For these reasons, the linear model is not discussed further in this work.
\subsubsection{Dirichlet prior}
The Dirichlet nonparametric prior specifies that the fractional specific star formation rate for each time bin follows a Dirichlet distribution \citep{leja17,leja18}. A Dirichlet distribution describes $N$ parameters $x_n$ which are bounded between $0 < x_n < 1$ and obey the constraint $\sum\limits_N x_n = 1$. Due to the summation constraint, only $N-1$ parameters are necessary to specify a Dirichlet distribution with $N$ bins. The $N^{th}$ model parameter is the logarithm of the total mass formed, which effectively controls the normalization of the SED. This cleanly separates the normalization and shape of the SFH.

The Dirichlet parameters $x_n$ are related to the fractional mass formed in each bin via:
\begin{equation}
m_n = \frac{x_n t_n}{\sum\limits_N x_n t_n}
\end{equation}
where $t_n$ is the width of each bin in time. The width of each time bin is included to center the prior on a constant SFR(t) = M$_{\mathrm{formed}}$/t$_{\mathrm{universe}}$, in contrast to the logM prior which is centered on a constant amount of mass formed in each bin. The method from \citet{betancourt12} is used to sample efficiently from a Dirichlet distribution. 

The Dirichlet distribution additionally requires a concentration parameter, \alphad, which controls the preference to put all of the weight in one bin (\alphad $< 1$) versus distributing the weight evenly between all bins (\alphad $\geq 1$). For this work a symmetric Dirichlet distribution is assumed, i.e. the concentration parameter has the same value for all bins. We test forms with \alphad = [0.2, 1]. The practical effect of the concentration parameter is clear in Figure \ref{fig:sfh_prior}: a low concentration weights toward bursty SFHs, while a higher concentration weights toward smooth SFHs.
\label{sec:continuity}
\begin{figure*}[t!h!]
\begin{center}
\includegraphics[width=0.95\linewidth]{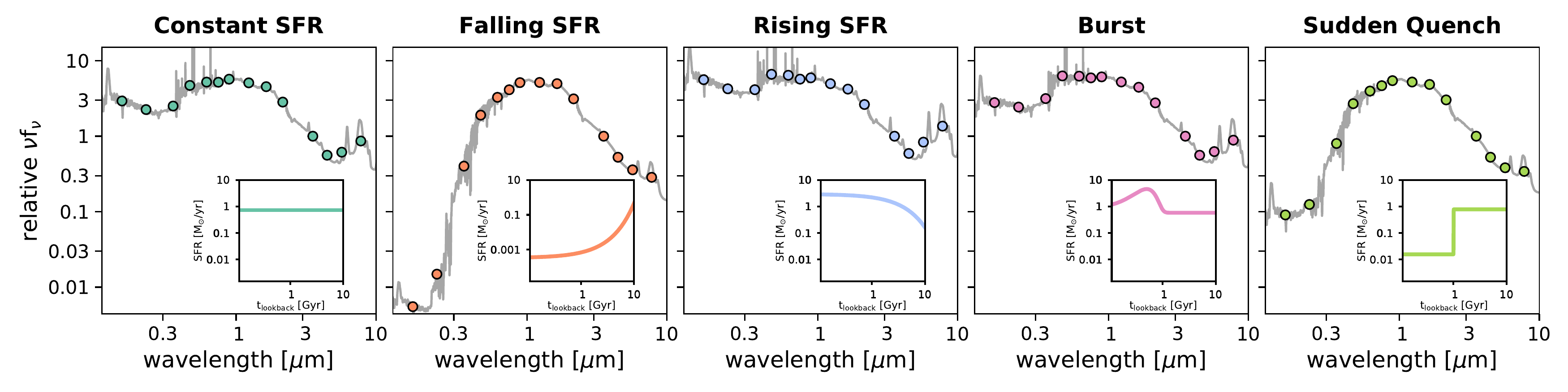}
\caption{Input star formation histories and resulting photometry for the mock tests. From blue to red wavelengths, the photometric bands are $GALEX$ FUV/NUV, SDSS $ugriz$, 2MASS $JHKs$, and $Spitzer$/IRAC channels 1-4.}
\label{fig:mock_phot}
\end{center}
\end{figure*}

The Dirichlet prior has several useful properties. Unlike the logM prior, the expectation value for SFR(t) is constant instead of rising. This results in a symmetric prior on stellar age and specific star formation rate. However, the Dirichlet prior does not enforce continuity in adjacent bins and thus permits sharp quenching and rejuvenation events, particularly at low concentrations. Additionally, it is not straightforward to add in additional physics as the only flexibility in the standard Dirichlet distribution is in tuning \alphad{}. Generalized Dirichlet functions exist such that SFR continuity can be enforced within the Dirichlet model, but these are mathematically challenging to work with and beyond the scope of this paper. 
\subsubsection{Continuity}
The continuity prior fits directly for $\Delta$log(SFR) between adjacent time bins. This prior explicitly weights against sharp transitions in SFR(t) and is similar to the regularization schemes used in STECMAP \citep{ocvirk06} and VESPA \citep{tojeiro07}.

Here the Student's-t distribution is adopted for the prior on $x=\log({\mathrm{SFR}_n / \mathrm{SFR}_{n+1}})$: 
\begin{equation}
\mathrm{PDF}(x,\nu) = \frac{\Gamma(\frac{\nu+1}{2})}{\sqrt{\nu\pi} \, \Gamma(\frac{1}{2}\nu)}\left(1+\frac{(x/\sigma)^2}{\nu}\right)^{-\frac{\nu+1}{2}}
\end{equation}
where $\Gamma$ is the Gamma function, $\sigma$ is a scale factor controlling the width of the distribution, and $\nu$ is the degrees of freedom controlling the probability in the tails of the distribution. Here $n$ ranges from 1 to $N-1$, and the $N^{th}$ variable is the total mass formed. The Student's-t distribution is chosen because it has heavier tails than the normal distribution, effectively allotting more probability to sharp transitions in SFR(t) such as quenching; note that as $\nu$ goes to infinity, the Student's-t distribution reverts to a normal distribution. We adopt $\nu=2$ and $\sigma=0.3$. Appendix \ref{sec:illustris} compares the chosen prior for $\log({\mathrm{SFR}(t) / \mathrm{SFR}(t+\Delta t)})$ against a distribution of this quantity measured from realistic star formation histories taken from the Illustris hydrodynamical simulation \citep{vogelsberger14a,vogelsberger14b,torrey14}. The choice of $\nu=2$ is similar to, but slightly wider, than the simulated distribution across a variety of redshifts and distributions of $\Delta t$, thus roughly matching the Illustris simulations while also allowing a reasonable amount of extra flexibility.

The continuity prior preserves many of the useful properties of the Dirichlet prior, including a symmetric prior in age and sSFR and an expectation value of constant SFR(t). The key difference from the Dirichlet prior is that the continuity prior explicitly weights against sharp changes in SFR(t). The choice of a Student's-t distribution ensures that the continuity prior remains flexible enough to describe maximally old and highly star-forming galaxies. Furthermore, the parameterization makes it straightforward to specify additional physics in the prior: for example, setting the mean of the distribution to mimic the cosmic star formation rate density as a function of time would weight galaxy SFHs towards the cosmic mean (e.g., \citealt{gladders13}). \footnote{Another practical upside to this model is that the run-time scales well with $N$: over a range of $4 \leq N \leq 13$, the run-time is essentially flat. This is likely due to the ease with which a nested sampling routine can explore Gaussian-like priors.}

\subsubsection{Flexible time bins}
\label{sec:flexprior}
\begin{figure*}[t!h!]
\begin{center}
\includegraphics[width=0.95\linewidth]{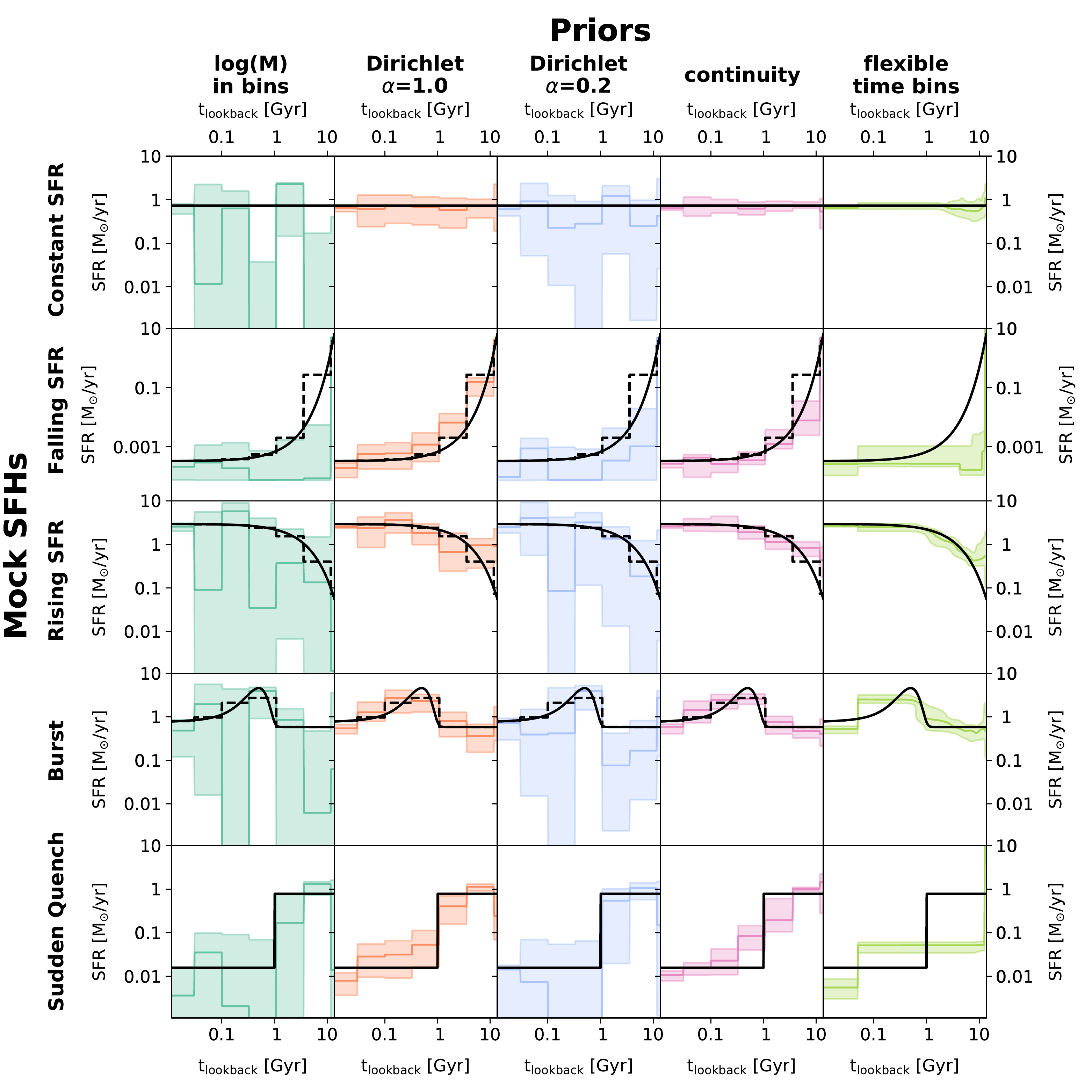}
\caption{Posterior SFHs from fitting mock photometry for different input SFHs and priors. Each row shows a different input SFH while each column shows a different prior. The input SFH is shown as a solid black line. The dashed black line is the input SFH re-binned to match the fixed time bins of the nonparametric models. The colored lines show the median of the posterior, while the shaded regions show the 16$^{\mathrm{th}}$-84$^{\mathrm{th}}$ percentiles of the posterior. The mock photometry has S/N=25.}
\label{fig:sfhs}
\end{center}
\end{figure*}
The priors described in the previous sections control how mass is distributed in fixed time bins. Here, we explore a model which instead distributes time for fixed mass fractions. This approach is similar to one described in \citet{iyer19}. 

This model splits the total mass formed into $N$ bins of equal mass $m$. The variable parameters are the edges of the time bins. As the first and last bins have two fixed edges ($t=0$ and $t=t_{\mathrm{univ}}$), there are $N-1$ remaining edges with the $N^{\mathrm{th}}$ parameter controlling the total mass formed. This method approaches a ``truly'' nonparametric model in that the only user inputs are $N$ (the number of time-resolving elements) and the prior, as opposed to the previous models which must also specify fixed time bins. The same continuity prior on $\log({\mathrm{SFR}_n / \mathrm{SFR}_{n+1}})$ is used to specify the prior for the distribution of time bin edges.

The primary downside of this method is that enforces a minimum sSFR$_{\mathrm{floor}}$(z) = ($N$t$_{\mathrm{univ}})^{-1}$. This floor is approached in the extreme where $N-1$ time elements have a infinitesimally small width (typically describing either a maximally old or maximally young population) and the final time bin spans the age of the universe. At $z=0$ with $N=7$ time elements, this corresponds to sSFR$_{\mathrm{floor}} \approx 10^{-11}$ yr$^{-1}$. This floor is a practical issue: sSFR = 10$^{-11}$ yr$^{-1}$ is a full order of magnitude higher than the sSFR of a typical quiescent galaxy at $z\sim0$ \citep{fumagalli14}. 

Accordingly, for this work we instead adopt a hybrid approach where there are two fixed time bins at $0 < t < 50$ Myr and $11.7 < t < 13.7$ Gyr which form masses $m_1$ and $m_2$, and the remaining mass $M-m_1-m_2$ is split between 5 flexible time bins. This allows arbitrarily low instantaneous specific star formation rates. The minimum allowed sSFR in the remaining flexible time bins is still present, but lowered by a factor of ($m_1+m_2/M$).
\begin{figure*}[t!h!]
\begin{center}
\includegraphics[width=0.95\linewidth]{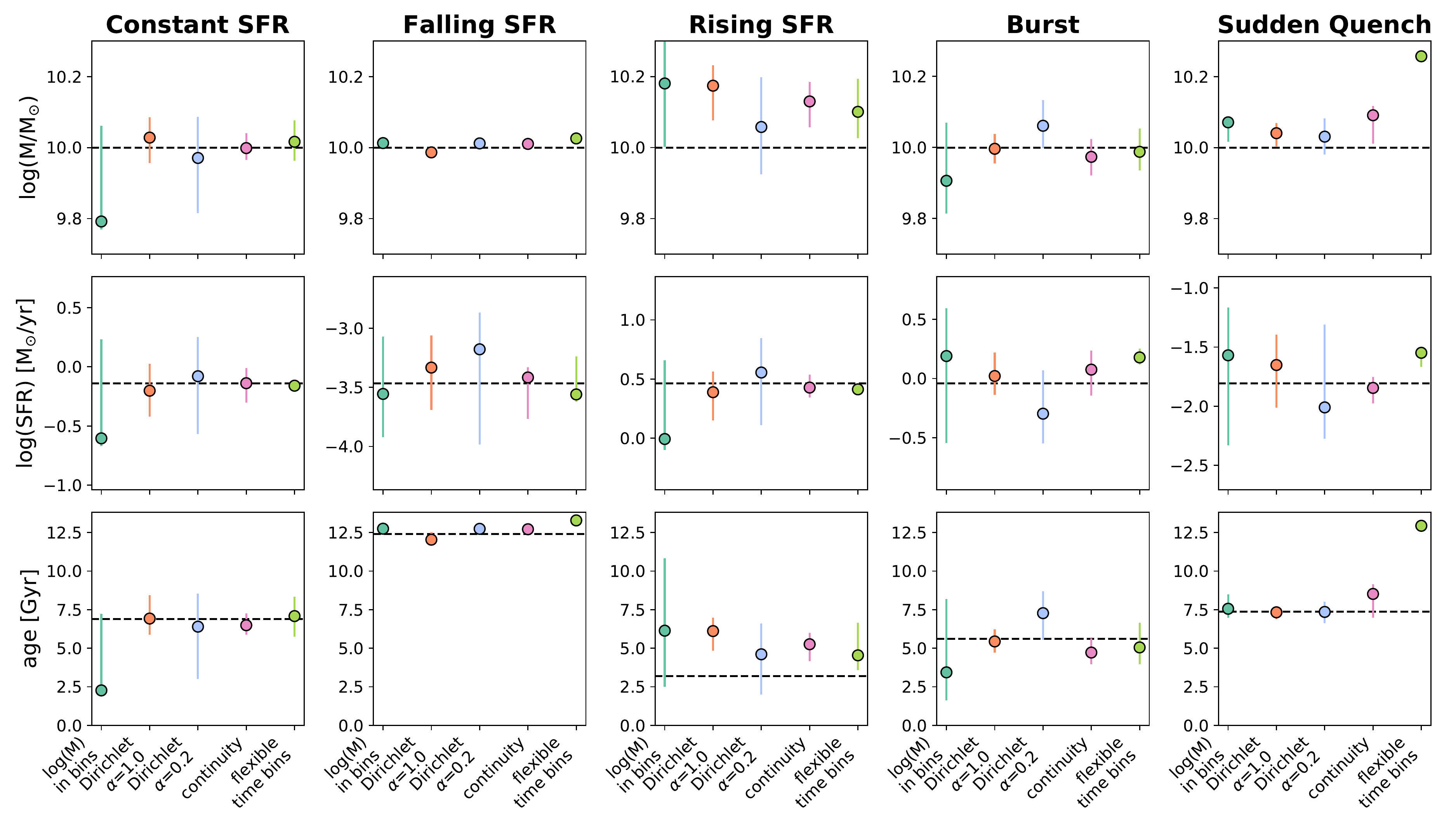}
\caption{Posterior masses, SFRs, and mass-weighted ages from fitting the mock photometry for a variety of input SFHs and adopted SFH priors. Each column shows the results from a different SFH prior, while each row shows a different parameter. The input values are shown with a dashed line. The SFR is calculated by averaging over the most recent 100 Myr.}
\label{fig:pars}
\end{center}
\end{figure*}

Overall, this approach retains the advantages of the continuity prior while also removing the discretization effects related to the fixed time bins. The tradeoff is the minimum allowed sSFR in the flexible time bins. This hinders the ability of the flexible time bins prior to accurately describe the recent and intermediate star formation history of quiescent galaxies, as shown in Section \ref{sec:sfhs}.

It is likely that additional modifications to this model can minimize or at least shift around the downsides. One potential alteration is to fit fixed bolometric luminosity fractions rather than fixed mass fractions. Due to the scaling of M/L with age, this alteration will remove the sSFR floor in the youngest time bins at the likely cost of imposing a new sSFR floor on the oldest time bins. Another possibility is to use bins with fixed but uneven mass fractions which can switch temporal ordering during the sampling phase. In principle, this can solve the sSFR floor by shifting low-mass bins to the youngest times when fitting the SEDs of quiescent galaxies. However, we do not explore such alterations further in this work.

\section{Generating and fitting mock galaxies}
\label{sec:mocks}
We generate and fit mock photometry with different nonparametric priors in order to explore the effect of the prior. To generate the mock photometry we use the Flexible Stellar Population Synthesis (\fsps{}) code \citep{conroy09b}. The mocks are generated with solar metallicity and a modest dust attenuation of A$_V = 0.3$. The following five star formation histories are used as inputs for the mocks:
\begin{description}
\item[Constant] SFR(t) = $M_{\mathrm{formed}}/t_{\mathrm{univ}}$ for all $t$.
\item[Declining] SFR(t) = $Ae^{-t/\tau}$, with $\tau$ = \tuniv/10 = 1.4 Gyr.
\item[Rising] SFR(t) = $Ae^{t/\tau}$, with $\tau$ = \tuniv/4 = 3.4 Gyr.
\item[Burst] A constant SFR forming 80\% of the total mass with a burst at $t=0.5$ Gyr forming 20\% of the total mass. The burst is a Gaussian with $\sigma$ = 200 Myr.
\item[Sudden Quench] A constant SFR until 1 Gyr in the past, where the SFR drops by a factor of 50.
\end{description}
These SFHs are chosen to represent the diversity of observed galaxy SFHs. All of the SFHs start at $t=0$ and extend to t$_{\mathrm{univ}}$ = $13.75$ Gyr. These SFHs are identical to the SFHs used in a companion paper \citep{carnall18b}.

The photometric bands cover far-UV (FUV) to mid-IR and include GALEX $FUV/NUV$, SDSS $ugriz$, 2MASS $JHK_s$, and $Spitzer$/IRAC channels 1-4. The input SFHs and resulting SEDs are shown in Figure \ref{fig:mock_phot}. Energy balance is adopted such that all energy attenuated by dust is re-emitted in the infrared. The fluxes are $not$ perturbed so as to avoid introducing random error into the results. The flux uncertainty used in the likelihood calculation corresponds to S/N values of 2, 5, 10, 25, and 100, applied uniformly across all bands. This test assumes homoscedastic noise and unperturbed fluxes, which simplify the analysis without loss of generality (discussed further in Section \ref{sec:sn_trends}). The effect of the prior when fitting real data with realistic noise properties such as heteroscedasticity and perturbations from the true measurements are explored later in Section \ref{sec:gama}.

\prospector{} \citep{prospector17,leja17} is used to fit the mock photometry. For each prior $N=7$ parameters are allotted to the star formation histories described in Section \ref{sec:priors}. The remaining free parameters are dust attenuation and stellar metallicity, with flat priors over $0 < \tau_V < 3$ and $-2 < $ log(Z/Z$_{\odot}) < 0.19$. 

The sampling is performed with the nested sampler \texttt{dynesty}\footnote{https://github.com/joshspeagle/dynesty} \citep{speagle19}. The \texttt{dynesty} package is a nested sampler which allocates live points dynamically, allowing the user to control the effective resolution as the fit proceeds. This allows the user to sample preferentially near the bulk of the posterior mass: for a fixed number of samples, this improves the estimate of the posterior at the cost of a higher relative error on the evidence. We take advantage of this by tuning the settings to emphasize accurate measurements of the posterior probability distribution: specifically, we require that the Kullback-Leibler divergence (KLD) between different realizations of the posteriors which include both statistical and sampling uncertainties to be less than 1.5\%. More details can be found in the \texttt{dynesty} documentation linked above.

\section{Mock results}
\label{sec:results}
Here we describe the effect of the SFH prior in fitting mock photometry.
\subsection{SFH recovery}
\label{sec:sfhs}
\begin{figure*}[t!h!]
\begin{center}
\includegraphics[width=0.95\linewidth]{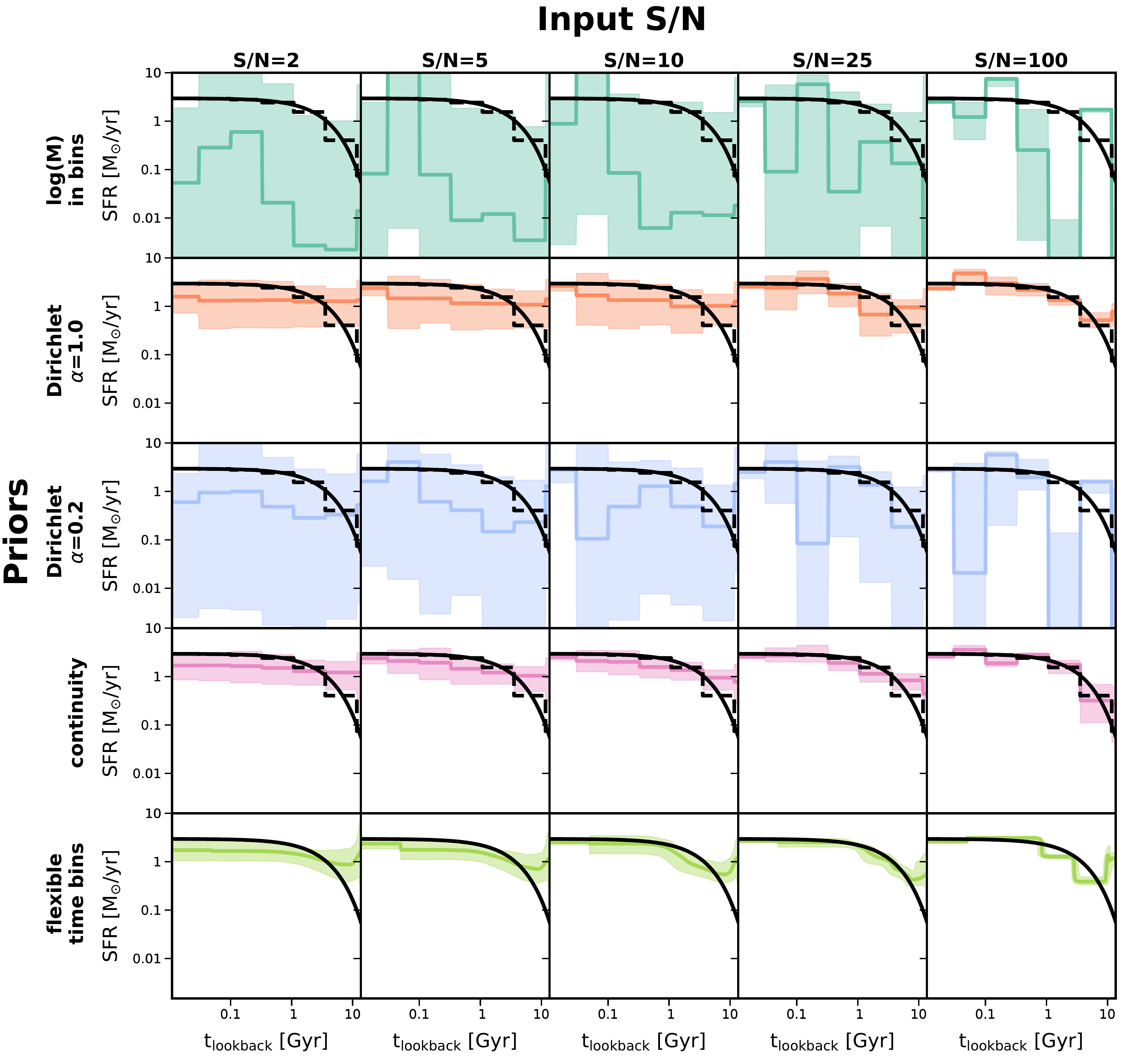}
\caption{The change in SFH posteriors caused by varying the input signal-to-noise ratio for an exponentially rising input SFH. The input S/N varies across the columns, while the priors vary down the rows. The input is a solid black line and the dashed black line is re-binned to match the nonparametric time bins. The median of the posterior is a thick colored line while the 16$^{\mathrm{th}}$-84$^{\mathrm{th}}$ percentiles are shown as shaded regions. While the SFH posteriors do shrink as the input S/N is increased, their shape and 1$\sigma$ ranges are primarily determined by the chosen prior rather than photometric S/N.}
\label{fig:sn_trends_sfh}
\end{center}
\end{figure*}
\begin{figure*}[t!h!]
\begin{center}
\includegraphics[width=0.8\linewidth]{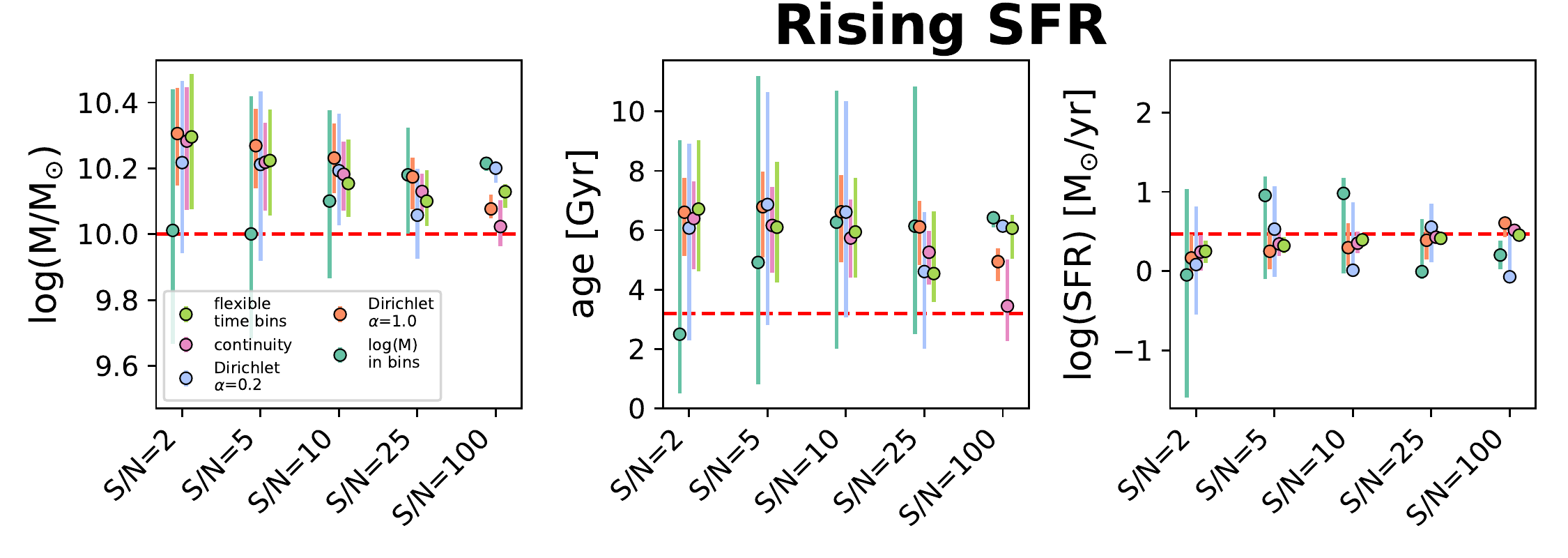}
\caption{The change in derived galaxy parameters while varying the signal-to-noise ratio for an exponentially rising input SFH. The three panels show the derived mass, age, and SFR as a function of prior and S/N. The derived parameters and their associated error bars tend to be dominated by the chosen prior rather than photometric S/N.}
\label{fig:sn_trends_pars}
\end{center}
\end{figure*}
\begin{figure*}[t!h!]
\begin{center}
\includegraphics[width=0.95\linewidth]{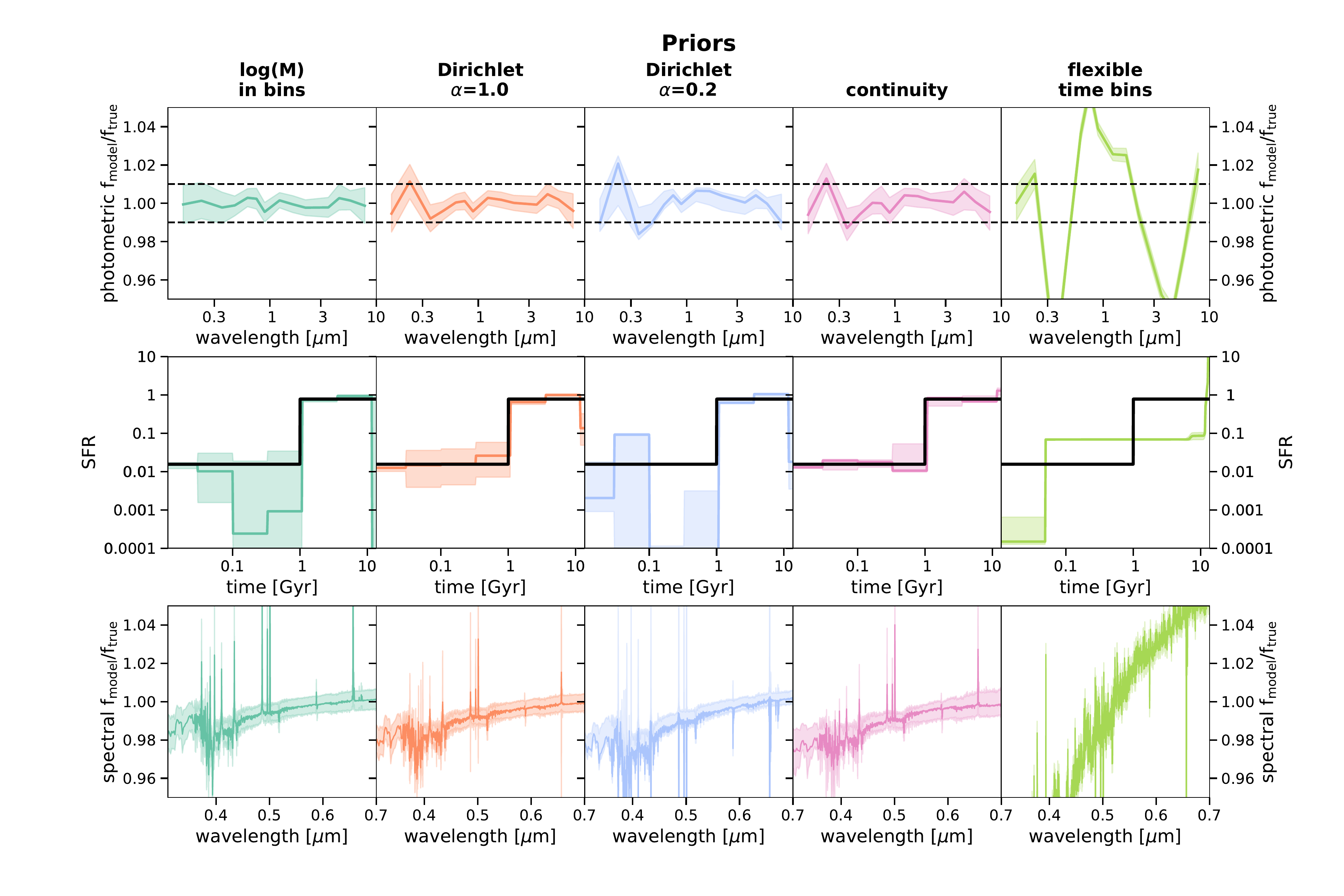}
\caption{Different priors can reproduce the S/N = 100 photometry to a similar level of accuracy while producing distinct posteriors. High-resolution spectra will be able to distinguish between these priors. The top row shows the ratio between the model posterior photometry and the true photometry. The 1$\sigma$ noise used in the likelihood calculation is shown with a dashed line. The middle row shows the input SFH in black compared to the posterior median (line) and 16$^{\mathrm{th}}$ and 84$^{\mathrm{th}}$ percentiles from fitting the photometry. While the model photometry is indistinguishable within the noise for most priors (excluding the flexible time bins prior), the star formation history posteriors remain distinct. The bottom row shows the same ratio, but for posterior model spectra. Absorption lines near the 4000$\AA$ break and strong emission lines such as H$\alpha$ and H$\beta$ can be used to further constrain the SFH posteriors after information in the photometry has been exhausted.}
\label{fig:evidence}
\end{center}
\end{figure*}
Figure \ref{fig:sfhs} compares the SFH posteriors to the input SFHs for all five nonparametric priors. This comparison is shown for mocks with a photometric S/N = 25. 

The shapes of the input star formation histories are fairly well-recovered for all priors. There is clearly sufficient information in the photometry to distinguish between rising, falling, and constant SFHs, and additionally some of the fits are able to recover the sharp quenching event and the recent burst of star formation. This demonstrates that nonparametric star formation histories can recover the zeroth-order behavior (rising, falling, constant) of the input SFHs regardless of the chosen SFR(t) prior.

In detail, however, there is significant variation in the location and size of the posteriors when different priors are applied. Indeed Figure \ref{fig:sfhs} clearly illustrates one of the main points of this study: even with high S/N photometry covering the far-UV-IR and no systematic errors, the posteriors are $strongly$ dependent on the prior. It is not uncommon for the SFR in fixed time bins to differ by $\gg1\sigma$ between priors, and the size of the $1\sigma$ range between priors are also quite different. For example, after fitting the constant input SFH with different priors, the 1$\sigma$ range for the posterior SFR in the $30 < t < 100$ Myr bin ranges from a factor of $\sim$100 for the logM prior to a factor of $\sim$2 for the continuity prior. This systematic difference in the width of the posteriors is consistent across a variety of input SFHs. These differences exist even though every nonparametric prior is flexible enough to accurately describe every input SFH\footnote{Excluding the combination of fitting the flexible time bins model to model SFHs with very low sSFRs; see discussion of these limitations in Section \ref{sec:flexprior}.}. This emphasizes the necessity of choosing a reasonable prior when fitting nonparametric SFHs and also the importance of understanding the influence of the chosen prior on the recovered parameters.

The priors can broadly be divided into two categories: priors which concentrate the majority of the mass in 1-2 bins (logM and Dirichlet $\alpha=0.2$) and priors which disperse the mass more evenly across all bins (continuity and Dirichlet $\alpha=1$). This division arises naturally due to the strong covariance between the SFR in nearby time bins: the data are often equally well reproduced by, for example, a fixed fraction of mass in old stars, regardless of whether these old stars formed steadily or in a single burst. The 'concentrated' priors are more successful at modeling input SFHs with sharp transitions such as steeply declining SFRs or sudden quenching events, whereas the 'dispersed' priors tend to focus the posterior mass on solutions with smooth SFR(t).

The differences in posterior star formation histories propagate directly into differences in derived physical parameters. Figure \ref{fig:pars} shows the recovery of mass, star formation rate, and mass-weighted age for each mock and prior combination at S/N = 25. Each of these parameters shows a different sensitivity to the prior. Mass is the most robust parameter, varying by $\sim$0.1 dex across each prior. Age and SFR are less robust, with the median of the posterior varying by up to 0.3-0.5 dex between priors. The prior also determines the size of the error bars. This is most dramatic for the age determination of the rising SFR, where the $1\sigma$ posterior for the logM prior spans 8 Gyr while the other $1\sigma$ posteriors range from 1-3 Gyr.

\subsection{Trends with S/N}
\label{sec:sn_trends}
Here we explore how the effect of the prior scales with the S/N of the mock photometry. Figure \ref{fig:sn_trends_sfh} shows the SFH posteriors for the rising SFH mock as a function of S/N, while Figure \ref{fig:sn_trends_pars} shows the same for the mock parameter recovery. Appendix \ref{sec:sn_append} shows the same results for the other four input SFHs.

These figures make it clear that the choice of prior is at least as important as the S/N in determining the posterior. We run a series of tests to quantify this comparison. We adopt the KLD as a distance metric; roughly this describes how far one probability distribution function (PDF) is away from another. For each combination of (S/N, SFH prior, input SFH), we measure the KLD between the posterior SFH and the posterior SFH of the fits with identical (SFH prior, input SFH) combinations but different values of S/N. We repeat the process while instead varying SFH prior. The KLD between posterior SFHs is taken as the sum of the KLDs of the marginalized PDFs in each SFH time bin. For the flexible time bin prior, we project the SFH posterior into the fixed time bins of the other SFH priors for this calculation.

This exercise confirms that choosing a different prior most often has a larger effect on the posterior than altering the input S/N. This is a remarkable finding: given that S/N scales with exposure time as S/N$\propto \sqrt{t}$, the range of S/Ns in this comparison corresponds to a 2500-fold difference in exposure time. This highlights the critical importance of choosing a prior. The falling SFH is the exception, where the information gain when going from S/N $\leq 5$ to S/N=100 is roughly equal to the effect of changing priors.

We note that adding perturbations to the fluxes and re-performing the fits would change the {\it median} of the derived posteriors in an unpredictable way. However, in general it will neither increase or decrease the overall {\it size} of the posteriors. Intuitively, the S/N of the data can be thought of as ``the distance which the model is allowed to stray from the best-fit solution'' which is, barring the case of extremely noisy likelihood surface, a weak function of the location of the best-fit solution. In this way, the posterior shape is fairly robust to the choice of adding perturbative noise. We note that we have verified this intuitive argument by comparing mock fits with and without perturbed noise.

These figures also demonstrate the posteriors do not converge even at S/N = 100. For example, the exact contribution of old stars in the $3.6 <t < 11.7$ Gyr time bin remains uncertain for all priors, resulting in a factor-of-two range in the posterior mass-weighted ages.

\subsection{Comparing residuals between priors}
The preceding analysis has demonstrated that different priors produce significantly different posteriors when fit to identical data. The last piece of the puzzle is to show the photometry is equally well reproduced by any of the adopted priors. 

Figure \ref{fig:evidence} shows the results of fitting S/N$=100$ photometry from the input SFH with a sharp quenching event. The posterior SFH and photometric residuals are shown for each prior. While each of these posteriors show a distinct pattern of behavior in SFR(t), the photometric residuals are largely within the 1$\sigma$ error range. Given that each model has the same number of parameters, the fact that each prior reproduces the data to a similar level of accuracy means that none of these priors are distinguishable with photometry alone\footnote{This has been confirmed directly by calculating the Bayesian model evidence, which suggests only a moderate preference at best for most combinations of prior and input SFH.}. The exception is the flexible time bins model, which is unable to accurately describe the input SFH because of limitations described in Section \ref{sec:flexprior}. This is the only such catastrophic failure amongst combinations of prior and input SFH.

The final panel of Figure \ref{fig:evidence} demonstrates that even after exhausting the information in the S/N=100 photometry, there is further constraining information available in high S/N spectroscopy. The predictions for the spectra after fitting the photometry are compared between different priors. Both the input spectrum and the posterior spectra are smoothed to a velocity resolution of $\sigma$=250 km/s. This suggests high S/N spectra can be fit simultaneously with photometry to further distinguish between nonparametric models. While calibration uncertainties make 2-3\% differences in the continuum very difficult to distinguish, even a simple constraint on the luminosity of strong nebular emission lines such as H$\alpha$ and H$\beta$ or break strengths such as the 4000 $\AA$ break can be helpful in ruling out some models.
\begin{figure*}[t!h!]
\begin{center}
\includegraphics[width=0.85\linewidth]{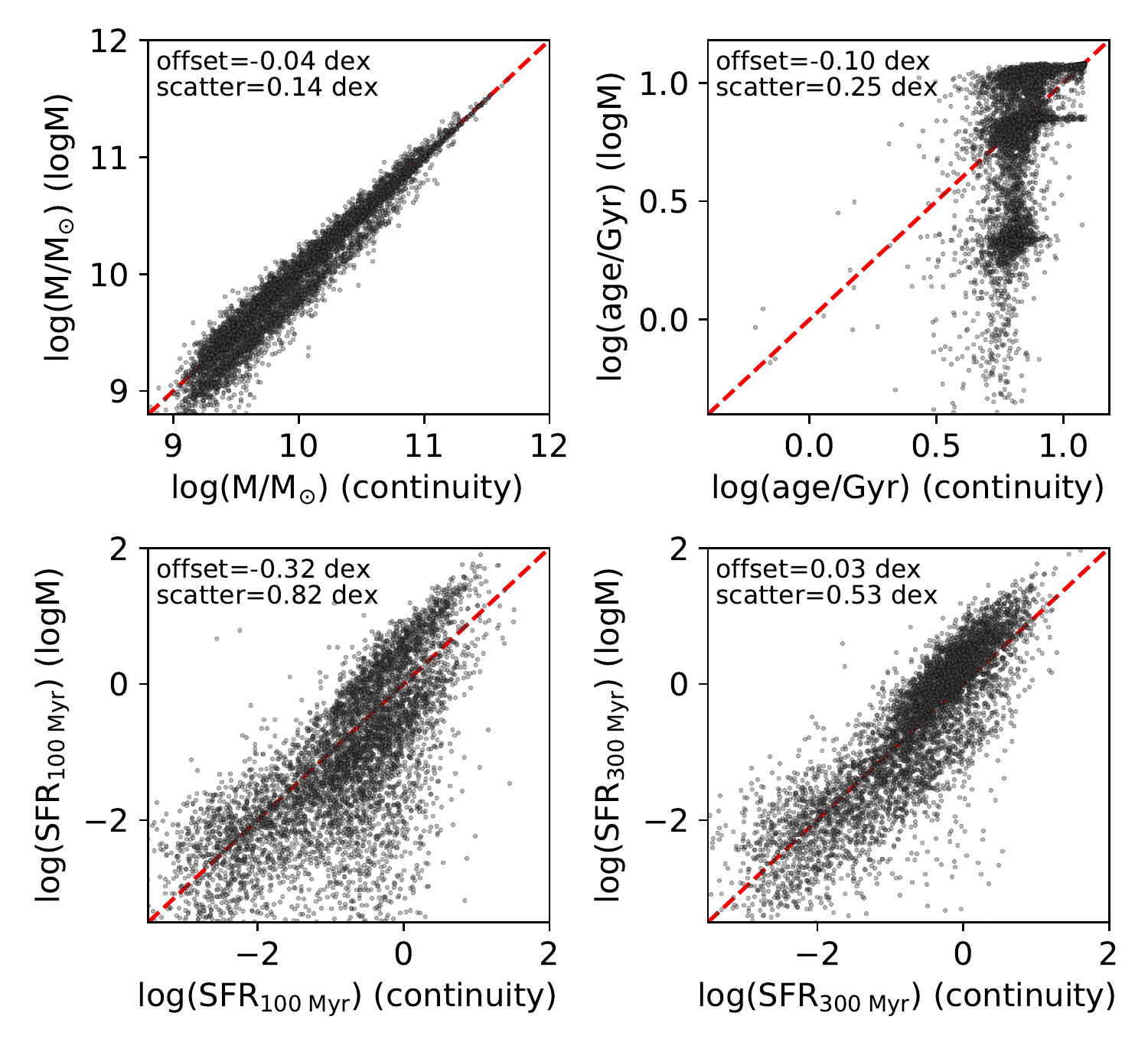}
\caption{The difference in derived parameters from SED fits with different SFH priors. The results from fitting the GAMA survey with both the logM and continuity priors are shown. The scatter between different priors can be thought of as the sensitivity to the permitted covariances in time between star formation history bins. The SFR is the most sensitive to this effect (particularly when averaged over shorter timescales) whereas the stellar mass is the least sensitive. The SFR on shorter timescales also has a bias in the log of the average depending on whether `bursty' SFHs (logM prior) or `smooth' SFHs (continuity prior) as preferred. The clustering of ages for the logM prior represents the center of the time bins. The star formation rates are in units of M$_{\odot}$/yr. The error bars in the lower right represent the median 1$\sigma$ error for each quantity.}
\label{fig:gama_dpars}
\end{center}
\end{figure*}
\section{Application to the GAMA survey}
\label{sec:gama}
Here we fit broadband photometry from the GAMA survey with a lightly modified version of the \mname{} model from \citet{leja17}, substituting in two different nonparametric SFH priors. The goal is to provide a practical demonstration of the effect of the prior on the output stellar populations parameters. The fit is performed with both the logM and the continuity priors described in Section \ref{sec:priors}. The same data are fit with parametric models in a companion paper (Carnall et al., submitted).
\subsection{Fitting UV-IR galaxy photometry from GAMA}
The photometry is taken from DR3 of the Galaxy and Mass Assembly (GAMA) survey \citep{driver11,baldry18}. The photometry is generated with the \textsc{LAMBDAR} code \citep{wright16} in three fields covering 180 degrees on the sky and includes 21 bands spanning the far-UV to the far-IR from $GALEX$, the SDSS, VISTA, $WISE$, and $Herschel$ \citep{driver16}. The GAMA spectroscopic redshifts are adopted as cosmological distance measurements. All galaxies in the redshift range $0.05 < z < 0.08$ with M$_*>10^{9}$ M$_{\odot}$ are fit, a total of 6134 galaxies. The stellar masses used for the selection are from \citet{taylor11} and the GAMA survey is mass-complete in this mass and redshift range.

The GAMA catalog uses the `forced photometry' technique where positional priors from high-resolution photometry are used to perform photometry on lower-resolution images. At the time of this writing, objects with negative flux values in a given passband are reported as zero in the catalog. In order to preserve the noise properties we manually replace these zeros with the last value measured in the iterative forced photometry algorithm from the GAMA catalogs. This last value in the iteration is the correct negative flux value. We also enforce a minimum error of 5\% in the photometry to reflect both underlying systematic errors in the models for stellar and dust emission and systematic effects in the measurement of the photometry.

In brief, \mname{} is a 15-parameter model. The main differences between this model and the model used in the mock tests are (a) more complex dust attenuation, (b) more flexible dust emission, and (c) a free gas-phase metallicity parameter. In more detail, \mname{} has seven parameters for the stellar mass and nonparametric star formation history, two parameters separately controlling the gas- and stellar-phase metallicity, three parameters to describe the shape of the IR SED and three parameters to describe the dust attenuation as a function of wavelength and stellar age. This model self-consistently uses the stellar ionizing continuum to power nebular line and continuum emission \citep{byler17} and applies energy balance to generate the IR emission. The stellar mass--stellar metallicity relationship measured from the SDSS is adopted as a prior \citep{gallazzi05}. We take the conservative approach of widening the confidence intervals from this relationship by a factor of two to account for potential unknown systematics or redshift evolution. As before, we use \texttt{dynesty} to sample the model posteriors.

The photometry is fit with the full \mname{} model, once with the logM prior and once with the continuity prior. Both priors reproduce the photometry to the same level of accuracy: the median $\chi^2$/N$_{\mathrm{phot}}$ values are 1.73 and 1.69 (mean: 2.42 and 2.36) for the continuity and logM priors, respectively. The slight improvement in $\chi^2$ for the logM prior is expected, as it is overall a more permissive prior for SFR(t). However, there are significant differences in the derived parameters between the two priors and it is shown below that the logM prior is most likely producing considerably less reliable outputs.

\subsection{Derived physical parameters}
\label{sec:gama_dpars}
The SFH prior can affect the resulting galaxy physical parameters in highly nontrivial ways. 

Figure \ref{fig:gama_dpars} shows the difference between the two priors for the derived stellar masses, mass-weighted ages, and star formation rates averaged over two different timescales, 100 Myr and 300 Myr. The scatter in this comparison can be thought of as the sensitivity of the derived parameter to the SFH prior, which sets the covariance between SFH time bins. The star formation rate averaged over 100 Myr is the most sensitive to the prior while stellar mass is the least sensitive. We have verified separately that the parameters from the continuity prior show less scatter when compared to parametric fits (though with non-negligible offset); the continuity prior can thus be thought of as the 'fiducial' model in this comparison, in the sense that behaves most like the standard parametric methods used in SED fitting.

In general, the prior plays a strong role in determining a parameter when the parameter either has a relatively small effect on the observable data or has a strong degeneracy with other parameters. For all of the quantities in Figure \ref{fig:gama_dpars}, the effect of the prior is maximized specifically for blue star-forming galaxies. This is caused by two distinct effects. 

The first is ``outshining'' \citep{papovich01,maraston10,pforr12,conroy13a}, whereby it is difficult to distinguish the presence of old, dim stars behind luminous young blue stars. In this regime, the prior outweighs the minimal information available in the photometry about the older stars. Outshining is responsible for much of the difference in derived ages: the logM prior prefers relatively young ages for blue star-forming galaxies while the continuity prior assumes a much more extended history of star formation. This bifurcation in ages changes the mass-to-light ratio, causing a secondary sequence below the 1:1 line in the stellar mass comparison where the continuity prior assigns relatively larger masses to these blue star-forming galaxies. 

The second effect is a degeneracy between young- and intermediate-age stars. This is typically the cause of discrepancies for galaxies that show a $\gtrsim 0.5$ dex difference in SFR(100 Myr) in Figure \ref{fig:gama_dpars}. Star formation in the younger time bins can instead be mimicked by star formation in the 100-300 Myr time bin followed by sharp cessation of star formation between 0-100 Myr. Star formation rates averaged over a 100 Myr timescale are thus more sensitive to the adopted SFH prior than SFRs averaged over a 300 Myr timescale. This effect also explains the bias between different priors in SFR(100 Myr) which disappears in SFR(300 Myr): while it is possible to reproduce the colors of blue star-forming galaxies by invoking a peculiar SFH with an excess of stars aged $100-300$ Myr and no star formation between $0-100$ Myr, stars older than $\sim$ 300 Myr are too red to perform the same trick.

Stellar ages show relatively lower scatter (0.25 dex) than star formation rates but are similar in that they can change significantly when very bursty SFHs are allowed. There is also clear artificial structure in the age scatter. This clustering is caused by the logM prior's preference to stack the majority of the mass in 1-2 bins (Section \ref{sec:logm_prior}) which naturally results in strong quantization of ages.
\begin{figure*}[t!h!]
\begin{center}
\includegraphics[width=0.8\linewidth]{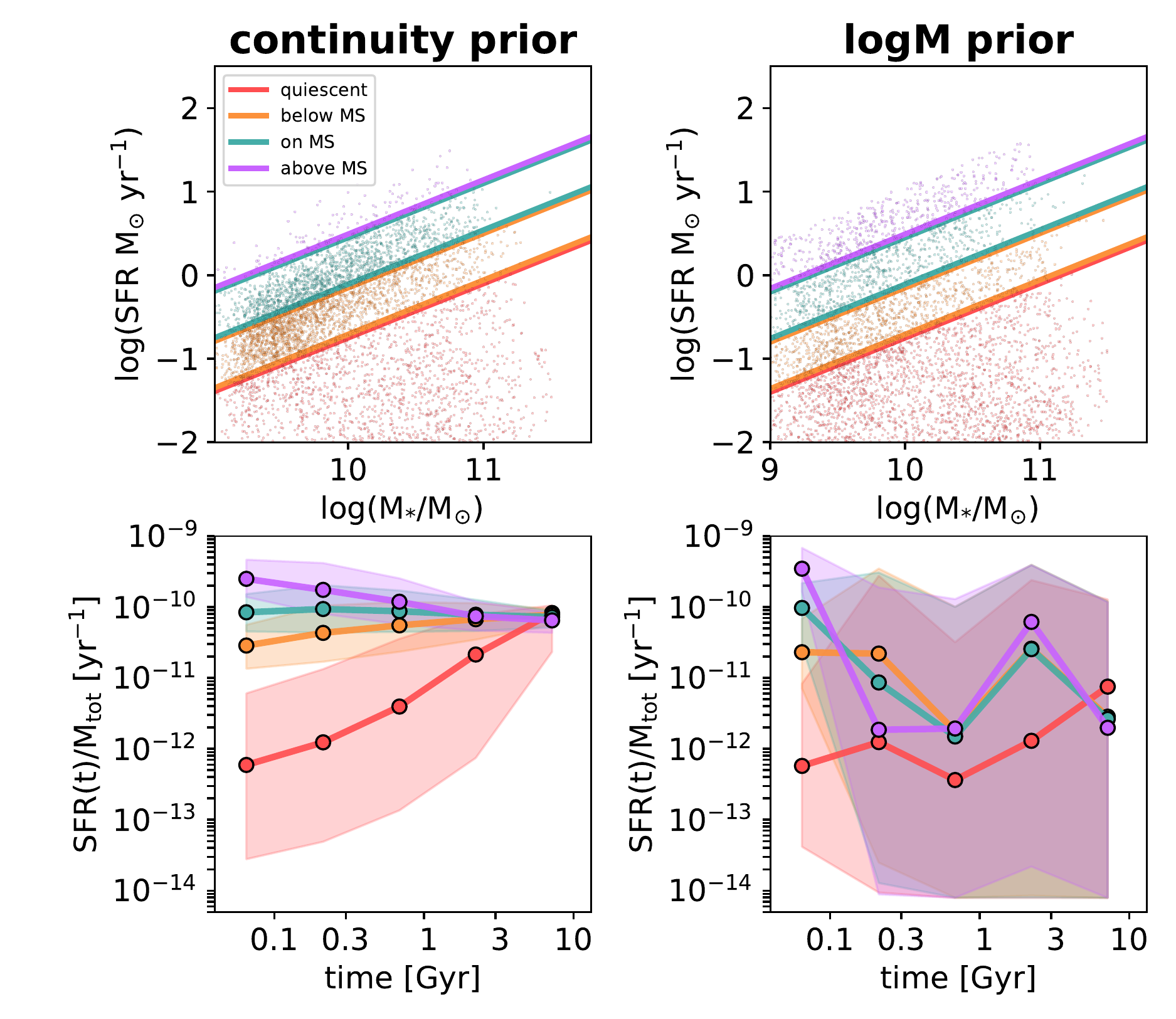}
\caption{Stacked SFHs across the star-forming sequence taken from fits to the GAMA photometry using two different SFH priors. The upper panels show the GAMA galaxies on the star-forming sequence with the locus of the star-forming sequence is taken from \citet{salim07}.  The SFH posteriors for each "slice" of the star-forming sequence (on, above, below, and quiescent) are summed and shown in the bottom panels. The median SFH is a solid line and the 16$^{\mathrm{th}}-84^{\mathrm{th}}$ percentiles are shaded regions. The two priors imply strikingly different galaxy assembly histories: for example, galaxies above the star-forming sequence have been above the star-forming sequence for $\sim1$ Gyr with the continuity prior or $\sim 30$ Myr (a lower limit imposed by the size of the youngest time bin) with the logM prior.
}
\label{fig:gama_sfh}
\end{center}
\end{figure*}
\begin{figure*}[t!h!]
\begin{center}
\includegraphics[width=0.8\linewidth]{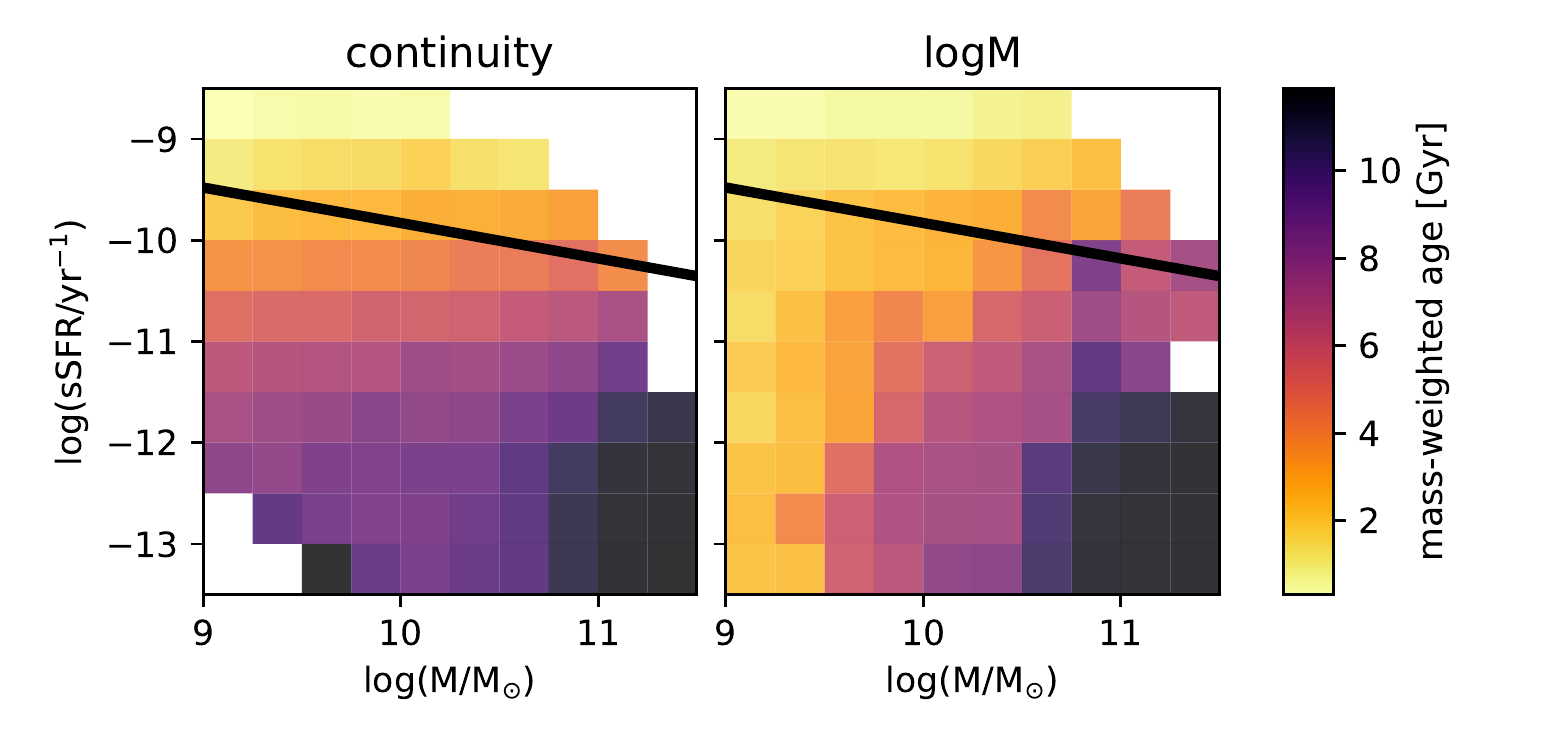}
\caption{The median of the stacked PDF over individual galaxies for mass-weighted age across the sSFR-mass plane. The left panel shows the results from the continuity prior while the right panel shows the results from the logM prior. The star-forming sequence from \citet{salim07} is marked in black. Uncolored cubes indicate that no galaxies fall into this region. The logM prior results in a larger range of ages and a more extreme trend between both age--mass and age--SFR. The logM prior also shows more scatter in these trends.
}
\label{fig:gama_cubes}
\end{center}
\end{figure*}
\subsection{Derived star formation histories}
\label{sec:gama_sfh}
It is further instructive to directly examine the difference in SFR(t) posteriors between the two priors.

In Figure \ref{fig:gama_sfh} the posterior star formation histories are stacked as a function of position on the star-forming sequence. The upper panels show the SFR measured in the most recent 100 Myr as a function of stellar mass for both priors. Colored lines divide the sample into four categories based on the location of the star-forming sequence. The locus of the star-forming sequence is taken from \citet{salim07} and the width is taken to be $+/-$0.3 dex \citep{speagle14}. This 0.3 dex width is used to distinguish between galaxies in four categories: quiescent, below the star-forming sequence, on the star-forming sequence, and above the star-forming sequence. The SFH posteriors of all galaxies in each category are stacked. The lower panels show the median SFR(t) and the +/- 1$\sigma$ posterior ranges for each stack. This can essentially be thought of as the range of SFHs displayed by galaxies inside of the specified mass and SFR selection. The posterior ranges include both measurement uncertainty and intrinsic scatter in galaxy SFHs. The two youngest SFH time bins (0-30 Myr, 30-100 Myr) have been combined in this plot as SFR(100 Myr) is used to select galaxies relative to the star-forming sequence.

The first result to note in Fig. \ref{fig:gama_sfh} is that there appears to be no star-forming sequence at all when fitting the data with the logM prior. This is quite striking as the star-forming sequence is found in many studies which fit broadband SEDs \citep{speagle14}. Here not only does assuming a logM prior fail to reproduce the star-forming sequence, but indeed it does so while providing a {\it better fit to the photometry} than the model which does reproduce the star-forming sequence. This means that the star-forming sequence is not a model-independent result when fitting photometry. Instead, it is contingent on assumption of smooth star formation histories which are typically built into SFH models.

Proving or disproving this assumption of smoothness in SFR(t) in the real universe is beyond the scope of this paper, though the preponderance of independent evidence does suggest smoother SFR(t) solutions than preferred by the logM prior. The current smoothness of SFR(t) implied by simulations and future prospects for testing this with observations are discussed in Section \ref{sec:sim_sfhs}.

Fits that assume a logM prior also infer a considerably in a considerably more chaotic process of galaxy formation than the continuity prior. For example, from Figure \ref{fig:gama_sfh}, the characteristic time a galaxy spends above or below the star-forming sequence is much shorter for the logM prior ($\sim 100$ Myr) than the continuity prior ($\sim$500 Myr). Fits using the logM prior imply that the typical galaxy above the star-forming sequence had an sSFR $\sim 10^{-12}$ yr$^{-1}$ approximately 200 Myr in the past. This suggests a rapid rejuvenation-quenching cycle for star-forming galaxies which is inconsistent with most numerical models of galaxy formation. The quenching timescales for quiescent galaxies are also sensitive to the prior: the logM prior quenches most galaxies much earlier than the continuity prior, but then also implies that $\sim20$\% of the quenched population was $above$ the star-forming sequence just $200$ Myr earlier. To be clear, this behavior is caused by the choice of prior: whether or not this behavior actually occurs in these galaxies is contingent on the accuracy of the prior.

Another dimension of this systematic prior difference is shown in Figure \ref{fig:gama_cubes}. This shows the median of the stacked PDF for mass-weighted age for individual galaxies across the SFR-mass plane. The fits with the logM prior show a range of ages between $1-12$ Gyr with a strong correlation with mass and moderate correlation with SFR. The fits with the continuity prior show a range of ages from $3-12$ Gyr with weaker but smoother correlations with mass and SFR. The logM results suggest that galaxy evolution proceeds in leaps and bounds, with some galaxies of 10$^{10}$ solar masses assembling half of their mass over as short as $\sim 1$ Gyr, whereas the continuity prior suggests a more steady and gradual process of mass assembly and longer minimum half-mass assembly times of $\sim3-4$ Gyr. 

In summary, the sensitivity of the SFR(t) posterior to the adopted prior means that the GAMA photometry is consistent with strikingly different assembly histories for both star-forming and quiescent galaxies. While it is not clear from these results alone which prior (if either) is ``correct'', it is clear the prior is a key determinant of the result.

\section{Comparing parametric to nonparametric SFHs}
\label{sec:par_vs_nonpar}
The results of this work suggest that in many cases, nonparametric SFHs cannot be fully constrained by typical galaxy observations. In spite of this -- indeed, in part because of this -- they offer several critical advantages over parametric models.

\subsection{Nonparametric SFHs Are More Flexible}
Nonparametric SFHs are flexible enough to describe the diversity of SFH shapes seen in galaxy formation simulations \citep{simha14,diemer17}, though the accuracy is limited by the resolution of the adopted time bins. This is in contrast to simple parametric models. For example, exponentially declining $\tau$-models are unable to produce the rising star formation histories seen in high-redshift galaxies \citep{maraston10,papovich11}. A delayed-exponentially declining model avoids this complication but still directly couples the shape of the short- and long-term SFH \citep{simha14}. More flexible parametric models such as double power-law or lognormal SFHs allow both rising and falling components but remain unable to model bursts and other sharp transitions in SFR(t). More generally, as the number of parameters in parametric fit becomes large, the lines between parametric \footnote{Parametric being defined here as belonging to a particular parametric family of probability distributions. For example, under this definition a spline would be nonparametric.} and so-called ``nonparametric'' fits will blur. In this scenario it remains true that a ÒnonparametricÓ approaches are superior because for such an approach the prior can be specified to arbitrary precision, as opposed to using a functional form in which the exact nature of the equation will influence the prior at some level. In this way, nonparametric models are the best family of solutions in which to fit arbitrarily complex behavior with time.

\begin{figure*}[t!h!]
\begin{center}
\includegraphics[width=0.95\linewidth]{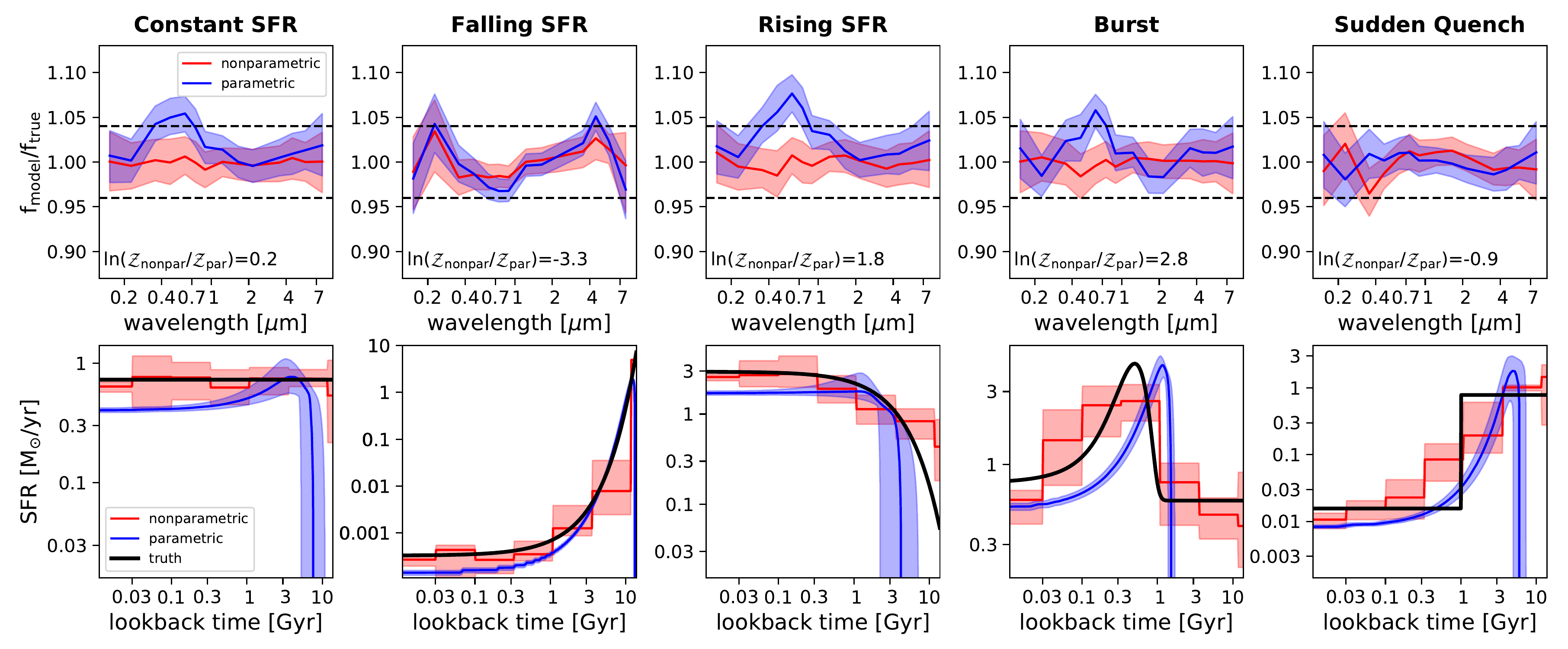}
\caption{Comparing photometric residuals and derived star formation histories from parametric and nonparametric fits to mock photometry. The parametric fit uses a delayed-tau model while the nonparametric fit uses the continuity prior. The upper row of panels show the posterior photometric residuals as a function of wavelength for each mock galaxy; the colored lines are the median while the shaded region is the 16$^{\mathrm{th}}$-84$^{\mathrm{th}}$ percentile. The dashed lines indicate the 1$\sigma$ photometric noise. The lower row of panels shows the posterior SFHs for each mock galaxy with the same color coding. The nonparametric fit both shows less bias and also returns more accurate error estimates than the parametric fit even though the Bayesian evidence doesn't consistently prefer either model.}
\label{fig:par_nonpar_comp}
\end{center}
\end{figure*}

This effect is illustrated in Figure \ref{fig:par_nonpar_comp}, which contrasts parametric and nonparametric fits to the mock galaxies from Section \ref{sec:mocks} with S/N = 25. The parametric SFH is a delayed-tau model,
$$
\mathrm{SFR}(t) = Ate^{-t/\tau}
$$
an SFH parameterization commonly used in the literature. The nonparametric model uses the continuity prior described in previous sections. 

The ratio of the Bayesian evidence in Figure \ref{fig:par_nonpar_comp} measures the relative evidence between two models. The evidence is conceptually similar to reduced $\chi^2$ in frequentist statistics: the evidence increases when the model can accurately reproduce the photometry, but is penalized for model complexity. Here, the Bayesian evidence doesn't strongly or consistently favor parametric or nonparametric fits to the input SFHs. The largest difference in evidence occurs for the falling input SFH, with $\sim$97\% odds (slightly more than 2$\sigma$) that the parametric fit is a better model -- not surprising, as the delayed $\tau$-model used to fit is very similar to the input $\tau$-model. Overall, if the goal is to simply reproduce the photometry from a diverse range of SFHs with a minimalist model, there is no clear preference between parametric and nonparametric models.

However, while the quality of fit to the photometry is largely similar, the accuracy of the recovered SFHs differ substantially. This is largely due to the relative rigidity of the parametric model. For example, while the parametric fit to the sudden burst mock galaxy does include a recent burst of star formation, the burst is older than expected by a factor of $\sim$3 and there are no stars older than $\sim$1.5 Gyr in the posterior. This happens because the structure of parametric SFH constrains it to {\it either} have a recent burst of star formation {\it or} have old stars. The compromise solution is to include a burst but one that is considerably older than the input. In contrast, the nonparametric model accurately describes the age of the burst and also predicts the correct mass in old stars. The fits to the other mock SFHs tell a similar story: the nonparametric model is flexible and can reasonably emulate any input shape whereas the parametric model must often be twisted or distorted to describe the input SFHs which necessarily inserts bias in the resulting SFH posterior.

In summary, the flexibility of nonparametric SFHs allows them to more accurately describe an arbitrary input SFH. This is important because any bias in the posterior SFH will be propagated into biases in derived galaxy parameters (see Figure \ref{fig:pars}). This means nonparametric SFH fits should recover galaxy parameters such as mass, age, and sSFR with less bias than parametric fits.

\subsection{Nonparametric SFHs allow explicit control over the prior density}
In addition to flexibility, nonparametric fits offer explicit control over the allowed density of SFR(t) models, i.e. the prior. Parametric fits typically impose highly informative priors on the shape of the galaxy star formation history. Such informative priors are not always intended but are necessary consequences of the chosen parametric form. These translate directly into informative priors on stellar age, stellar mass-to-light ratio, and specific star formation rate. The priors imposed by parametric models are shown directly in Carnall et al. (submitted). These informative priors rule out entire classes of SFR(t) solutions before the data are fit and can result in unrealistically tight posteriors. 

Conversely, in nonparametric models the priors can be directly tuned in order to capture the distribution of galaxy behaviors (see Figures \ref{fig:sfh_prior}, \ref{fig:par_prior}). This tuning is especially important when an SFH model is not well-constrained by the available data. This is a strong argument for the use of nonparametric models even when fitting data with little constraining power. In such cases it is important to marginalize over a full range of realistic SFR(t) behaviors in order to get realistic answers, rather than marginalizing over the (typically much smaller) SFR(t) solutions allowed by parametric models.

This contrasting behavior can be seen in Figure \ref{fig:par_nonpar_comp} where the parametric fits have much tighter posteriors than the nonparametric fits for all input SFHs. This is due to the (often unintentional) informative priors imposed by parametric models. These priors result in overly tight posteriors even when the input SFH is a good match to the parametric form, as is the case for the fit to the exponentially declining input SFH. This effect is explored in more detail in the companion paper by Carnall et al (submitted).

\begin{figure*}[t!h!]
\begin{center}
\includegraphics[width=0.8\linewidth]{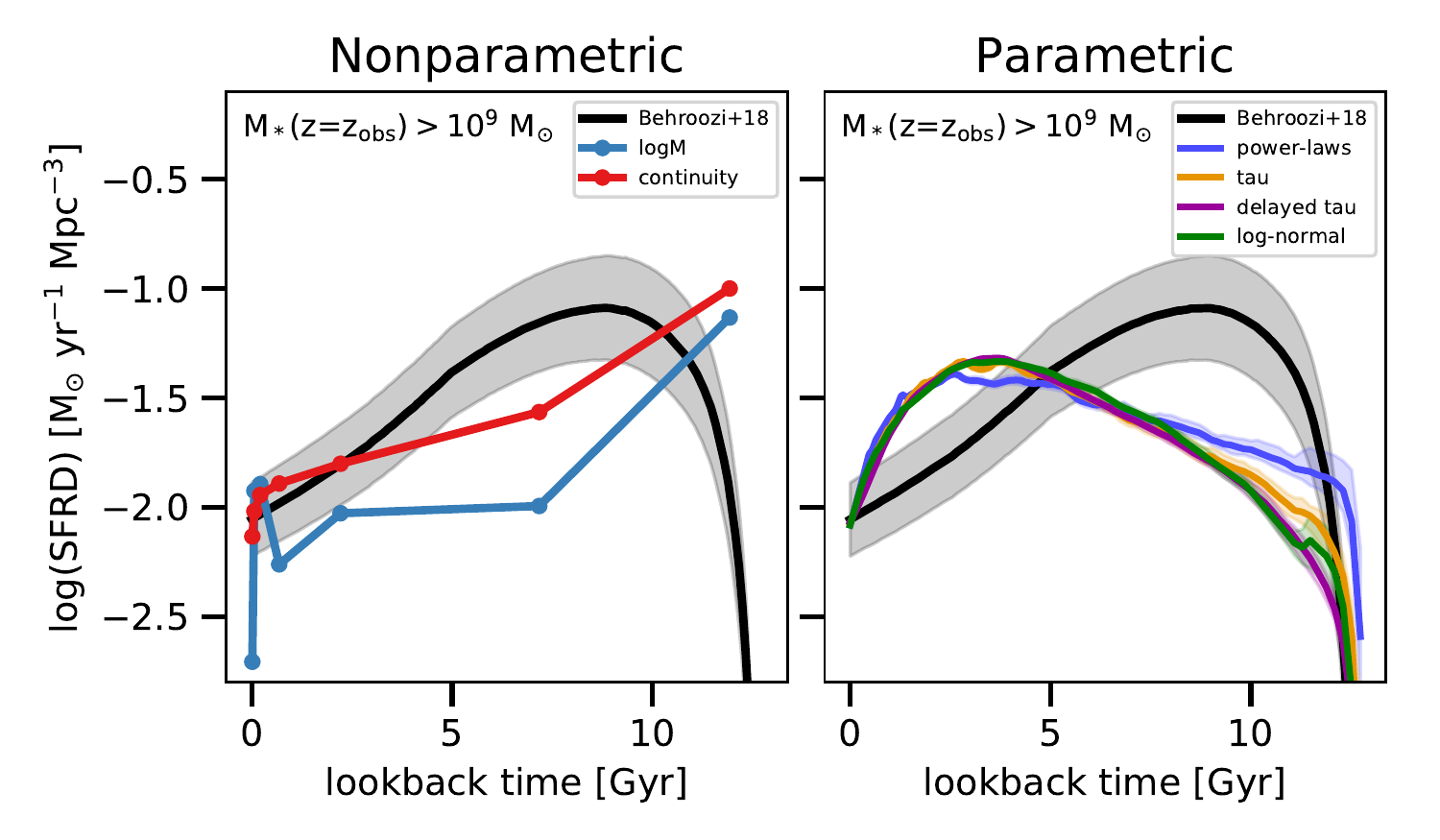}
\caption{The cosmic star formation rate density measured by fitting the GAMA sample with different SFH models. The left panel shows the nonparametric SFH models described in Section \ref{sec:gama} while the right panel shows parametric SFH models fit to the same data in a companion paper (Carnall et al., submitted). Empirical predictions from the Universe Machine are shown as a black line \citep{behroozi18}. Both the empirical predictions and the SFHs from GAMA include only galaxies with measured stellar mass $>10^{9}$ M$_{\odot}$. The SFHs from GAMA are normalized such that they have the same average value over the most recent 100 Myr as the Universe Machine; this is done to highlight differences in the shape of the derived CSFRD. Overall, parametric models underestimate t$_{\mathrm{half}}$, the time at which half of the cosmic mass budget had assembled, by $\sim$50\%, while nonparametric models overestimate t$_{\mathrm{half}}$ by a more modest $\sim$25\%.}
\label{fig:sfrd}
\end{center}
\end{figure*}

Another example is in Figure \ref{fig:sfrd}, which shows cosmic star formation rate density (CSFRD) measured directly from the SFHs fit to the GAMA photometry (Section \ref{sec:gama}). In principle, the CSFRD observed at higher redshifts should be reproducible by the SFHs of a low-redshift galaxy sample (e.g. \citealt{heavens04}). Here we show the CSFRD for both parametric and nonparametric models, with the parametric fits taken from the companion paper by Carnall et al. (submitted). We compare to the cosmic star formation rate density from the Universe Machine \citep{behroozi18}. The empirical measurements from the Universe Machine include only galaxies with M$_*(z=0) \lesssim 10^{9}$ M$_{\odot}$ in order to match the selection function used in Section \ref{sec:gama}. Measurement error from the observed star formation histories is very low except for large lookback times in the parametric models. The Universe Machine error contours are from the systematic offset between observed and true stellar mass within the Universe Machine model.

Here it can be seen that parametric models systematically underestimate galaxy ages: indeed, parametric models underestimate t$_{\mathrm{half}}$, the time at which half of the cosmic mass budget had assembled, by $\sim$50\%, while nonparametric models overestimate t$_{\mathrm{half}}$ by a more modest $\sim$25\%. Since the oldest stars have the most subtle observational signature, the SFH at the oldest lookback-times is the most prior-dominated. The simplest explanation for this difference is that the time at which star formation begins is typically a free parameter in parametric models, while in these nonparametric models star formation always begins at $t=0$. This causes galaxy ages from parametric models to be systematically lower than nonparametric models, consistent with findings in the literature that parametric models significantly underestimate galaxy ages \citep{wuyts11a}. Accurate galaxy ages are important as the optical mass-to-light (M/L) ratio varies as M/L $\propto$ (age)$^{0.6-0.7}$ (e.g., \citealt{elmegreen12}), implying that parametric models underestimate galaxy masses by $\sim$40\% while nonparametric models overestimate it by a smaller $\sim15\%$. It is simple to adjust this in nonparametric models by tuning the mean SFR(t) directly, but the prior for SFR(t) in parametric models is difficult to tune as it is coupled to the assumed parametric form.

Ultimately the SFH prior will be important in almost all applications of SED modeling to galaxy photometry, as even at S/N $=100$ the mock tests in this work show that nonparametric models produce distinct posteriors in derived parameters and in SFR(t) (Figures \ref{fig:sn_trends_pars}, \ref{fig:sn_trends_sfh}, and Appendix \ref{sec:sn_append}). In practice, 1\% accuracy is very difficult to achieve due to systematic effects in measuring galaxy photometry such as background subtraction, light profile modeling, and contamination from nearby objects (e.g., \citealt{bernardi13,skelton14}). This suggests that studies which attempt to recover galaxy SFHs from photometry will almost never be tightly constrained by the available data, re-emphasizing the importance of having SFH models with easily customizable priors.

\section{Choosing the right prior: comparison to simulations}
\label{sec:sim_sfhs}
The previous sections have made the case that nonparametric models will never be fully constrained by typical galaxy photometry. Therefore, the prior will always have at least a moderate role in determining the answer. In light of this fact, it is critical to choose a well-motivated prior.

A well-motivated prior will be one which best mimics the breadth and relative distribution of SFR(t) in galaxies. One key question is therefore on what timescales SFR(t) changes in real galaxies. Theoretical models of galaxy formation suggest two potential answers: either this timescale is primarily set by processes related to the halo dynamical time (e.g., the gas accretion rate) as in the equilibrium bathtub model (e.g. \citealt{dave11,forbes14}) and cosmological hydrodynamical simulations \citep{crain15,schaye15,pillepich18} or it is set by the timescales of star formation feedback as in the high-resolution Feedback In Realistic Environments (FIRE) simulations \citep{hopkins14}.

\citet{torrey18} explore star formation timescales in the IllustrisTNG simulation. They find that these timescales scale with the halo dynamical time. They fit exponential decay curves of the form exp($-\delta t / \tau$), effectively identifying the shortest timescales on which there is significant variation in SFR. Over a range of $0 < z < 4$ and $9 < \log(\mathrm{M/M}_{\odot}) < 11$, they find values ranging from $0.2 \lesssim \tau \lesssim 3$ Gyr. Averaged over the galaxy population this timescale changes from 0.25 Gyr at $z=4$ to 1.5 Gyr at $z=0$, and also decreases with decreasing stellar mass.

In contrast, strong stellar feedback in the FIRE simulations produces globally-bursty star formation histories. In FIRE, the shortest bursts occur on timescales of 5-50 Myr, a factor of $\sim5-10$ shorter than the Illustris simulations (see Figure 9 in \citealt{sparre17}). This is a strong function of mass: at $0 < z < 0.4$, galaxies with M$_* < 10^{10}$ M$_{\odot}$ change their SFR by orders of magnitude over 200 Myr, while galaxies with M$_* > 10^{10}$ M$_{\odot}$ change their SFR by a factor of two or less. At $z=2$, even massive galaxies have short bursts in which their SFR changes by an order of magnitude in a 200 Myr timespan.

The dichotomy between timescales regulation by evolution of the halo \citep{torrey18} versus timescale regulation by feedback maps roughly onto the behavior of `concentrated' priors (logM, Dirichlet \alphad $=0.2$) and `dispersive' priors (Dirichlet \alphad $=1$, continuity) respectively. This comparison is important for nonparametric time bins which are smaller than a few characteristic timescales ($\sim 1$ Gyr for Illustris and $\sim 50-100$ Myr for FIRE) so as to not average over short-term fluctuations. It is well-established that in both models the timescales scale strongly with galaxy mass and with redshift. Thus a more concentrated prior is more appropriate for high redshift and/or low-mass (M$_* \lesssim 10^{10} \, \mathrm{M}_{\odot}$) galaxies while a more dispersive prior is more suitable for low redshift and/or high-mass galaxies. This is interesting in light of the fact that concentrated priors return much larger errors on ages, masses, and SFRs. This suggests that given a low- and high-mass galaxy with identical colors, the ideal SED fit should return substantially larger error bars for the low-mass galaxy.

In principle these timescales can be discriminated via observational signatures that trace SFR over different timescales, such as the H$\alpha$/UV ratio \citep{weisz11b,kauffmann14,shivaei15,smit16}. However, this comparison is complicated by natural galactic variation in dust geometry and composition, initial mass function, and the uncertainty in ionizing photon production efficiency, all of which affect the H$\alpha$/UV ratio \citep{shivaei15,sparre17}. Another possible discriminator is the existence of the much-debated fundamental mass--metallicity--SFR relationship, which can only form if SFRs are regulated over longer timescales \citep{torrey18}. Finally, Figure \ref{fig:evidence} suggests that combined modeling of high S/N photometry and high S/N, high-resolution spectroscopy can be used to constrain the typical timescale over which galaxy SFRs change.

Ultimately, the ideal star formation history prior would likely be a function of galaxy stellar mass and the age of the universe and, for models with fixed time bins, the widths of the time bins. To keep the results general we deliberately do not adopt such scaling in this work. On a practical note, using time bins which change based on the age of the universe does create an implicit dependence of the prior on cosmological time.

Regardless of the prior used, it is advised that this prior is clearly stated when describing galaxy SFHs recovered from data. Comparisons of recovered SFHs between studies need to take into account different assumed priors. Comparisons between observations and simulations should be aware of the priors adopted in the observational analysis as this will strongly affect this comparison when the data are good, and will dominate the comparison when the data are poor.

\section{Conclusion}
\label{sec:conclusion}
In this work we explore the effect of adopting different Bayesian priors when fitting nonparametric SFHs to photometry. A variety of nonparametric priors are tested, ranging from the most straightforward logM prior which fits for the mass formed in fixed time bins to a continuity prior which emphasizes smooth behavior in SFR(t). 

To test the different influences of these priors, we generate mock UV-IR photometry with simple input SFHs and attempt to recover them with the SED-fitting code \prospector{}. The key results of the mock tests are:
\begin{itemize}
\item All of the priors recover the shape of the input SFHs with reasonable accuracy.
\item However, priors impose different shapes on the posteriors. The priors can roughly be divided into `dispersive' priors which prefer to spread the mass evenly across time bins and `concentrated' priors which prefer to concentrate mass in 1-2 time bins.
\item The primary determinant of the size of the posteriors -- and therefore the size of the error bars -- is not the photometric noise, but instead the chosen prior. This is true for a range of signal-to-noise from 2 to 100, roughly equivalent to 2500-fold difference in exposure time.
\item Aside from a few combinations of prior and input SFHs, the photometry is equally well reproduced by any of the adopted priors.
\end{itemize}

In order to demonstrate these effects in practice, we fit UV-IR photometry for $\sim$6000 massive galaxies at $0.05 < z < 0.08$ from the GAMA survey using a lightly modified version of the \mname{} model from \citet{leja17}. This is done for both the logM and the continuity priors. We find:
\begin{itemize}
\item Stellar masses are relatively stable between priors, while ages and star formation rates show more significant dependence on the prior.
\item This dependence is maximized for blue star-forming galaxies, because of outshining effects and a degeneracy between young (0-100 Myr) and intermediate-age (100-300 Myr) stars.
\item The two priors result in significantly different star formation history posteriors. The logM prior suggests that galaxy formation proceeds in a chaotic fashion and over short timescales, whereas the continuity prior suggests smoother evolution over longer periods of time.
\end{itemize}

It is tempting to conclude that the dependence of nonparametric SFHs on the prior is a weakness of the approach. This is misleading, however; due to their lack of flexibility, parametric models impose stronger priors on SFR(t) than nonparametric models. Parametric models achieve tight posteriors by {\it a priori} ruling out many different forms of solutions, resulting in unrealistically precise answers even when fitting low S/N data. Perhaps surprisingly, this means it is better to use a well-tuned nonparametric model -- even when it {\it cannot be constrained by the data} -- as it will produce more meaningful error bars.

The quality of the error bars from a nonparametric analysis is dependent on choosing a reasonable prior. Here, we have contrasted the performance of several different priors in accurately recovering mock input SFHs. Ultimately the ideal prior is one which best mirrors the distribution of galaxy star formation histories in the real universe. This distribution has yet to be definitively measured in observations and or converged upon in galaxy formation simulations.

H$\alpha$ luminosities, UV and FIR photometry, and high-resolution, high S/N spectroscopy covering a statistical sample of galaxies across cosmic time can help to constrain the timescales on which galaxies change their SFRs. These data can then be used to construct a more ideal prior for SFHs. We leave a detailed exploration of these issues to future work.

\acknowledgements
We thank Benedikt Diemer, Sandro Tacchella, and Kartheik Iyer for thoughtful comments and discussion. We further thank Benedikt Diemer for providing SFHs from the Illustris hydrodynamical simulation. J.L. is supported by an NSF Astronomy and Astrophysics Postdoctoral Fellowship under award AST-1701487. The computations in this paper were run on the Odyssey cluster supported by the FAS Division of Science, Research Computing Group at Harvard University. This research made use of Astropy,\footnote{http://www.astropy.org} a community-developed core Python package for Astronomy \citep{astropy13, astropy18}.

GAMA is a joint European-Australasian project based around a spectroscopic campaign using the Anglo-Australian Telescope. The GAMA input catalogue is based on data taken from the Sloan Digital Sky Survey and the UKIRT Infrared Deep Sky Survey. Complementary imaging of the GAMA regions is being obtained by a number of independent survey programmes including GALEX MIS, VST KiDS, VISTA VIKING, WISE, Herschel-ATLAS, GMRT and ASKAP providing UV to radio coverage. GAMA is funded by the STFC (UK), the ARC (Australia), the AAO, and the participating institutions. The GAMA website is http://www.gama-survey.org/ .

\software{\texttt{Prospector} \citep{prospector17}, \texttt{python-fsps} \citep{pythonfsps14}, \texttt{Astropy} \citep{astropy13,astropy18}, \texttt{FSPS} \citep{conroy09b}, \texttt{matplotlib} \citep{matplotlib18}, \texttt{scipy} \citep{scipy}, \texttt{ipython} \citep{ipython}, \texttt{numpy} \citep{numpy}}

\appendix
\section{Testing the sensitivity to number and location of time bins}
\label{appendix:tbins}
Here we examine the sensitivity of the results to the number of time bins and their relative spacing. We show the results from the continuity prior for these tests, though the results are independent of the adopted prior. We fit the same mock data described in the main text, varying the number of bins between $4-14$. The data are assigned a signal-to-noise of 25 and the fluxes are not perturbed as per the fits in the main text.

The new bin spacing in time must be specified when changing the number of bins. In these tests the two youngest bins remain fixed at $0-30$ and $30-100$ Myr to minimize the allowed SFH variability on short timescales, which in turn avoids the instabilities in SFR estimates described in Sections \ref{sec:gama_dpars} and \ref{sec:gama_sfh}. Similarly, the oldest bin remains fixed at $11.7 < t < 13.7$ Gyr for the purpose of modeling a maximally old population. The remaining time between 0.1-11.7 Gyr is split into are split into equal logarithmic chunks and divided among the remaining bins. 

\begin{sidewaysfigure*}
\includegraphics[width=\textheight]{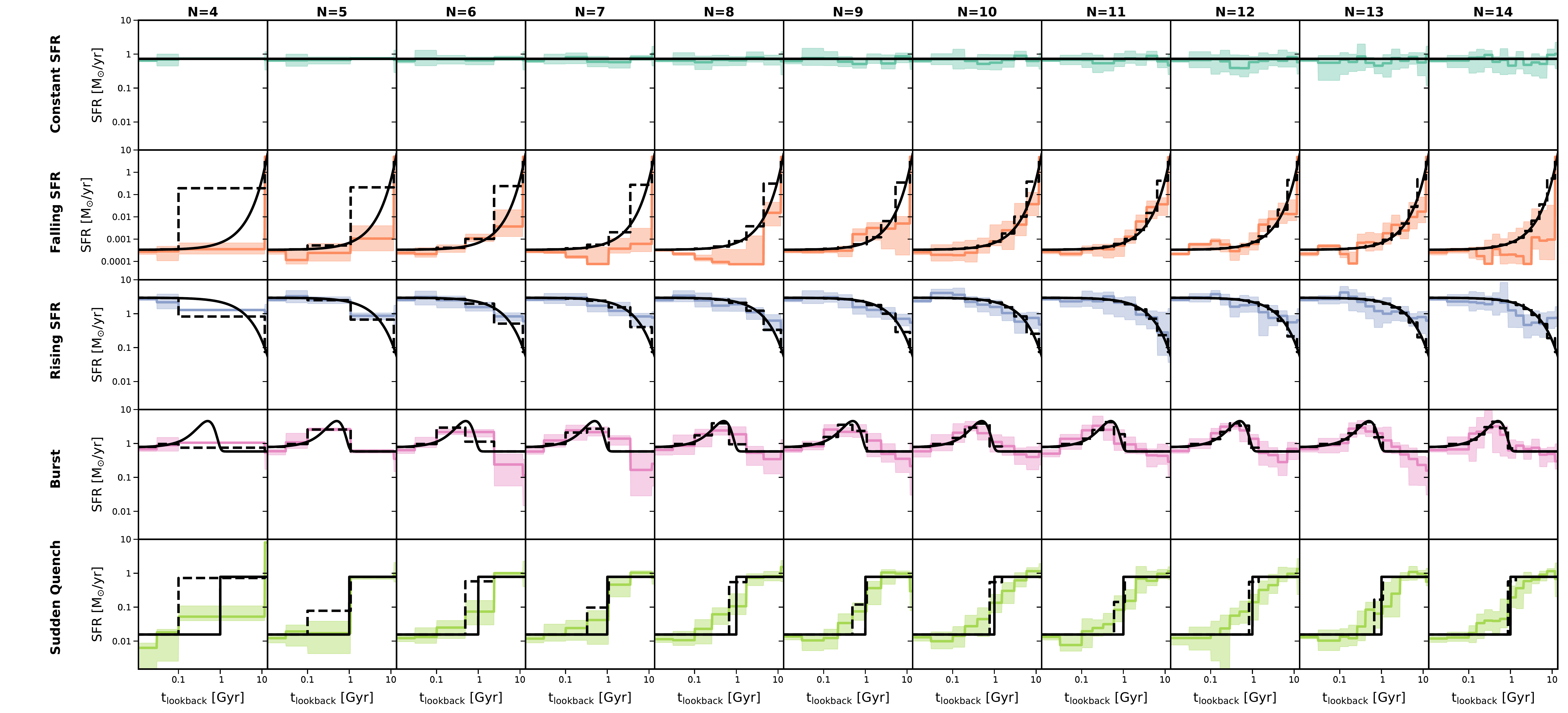}
\caption{Variation in posterior SFHs with N$_{\mathrm{bins}}$ derived by fitting mock photometry with different input SFHs. The posterior SFHs are largely independent of the number of SFH bins. Each row shows a different input SFH while each column shows a different number of SFH time bins. The input SFH is shown as a solid black line. The dashed black line is the input SFH re-binned to match the fixed time bins of the nonparametric models. The colored lines show the median of the posterior, while the shaded regions show the 16$^{\mathrm{th}}$-84$^{\mathrm{th}}$ percentiles of the posterior. The mock photometry has S/N=25.}
\label{fig:mock_timebin_sfh}
\end{sidewaysfigure*}

\begin{figure*}[t!h!]
\begin{center}
\includegraphics[width=0.85\linewidth]{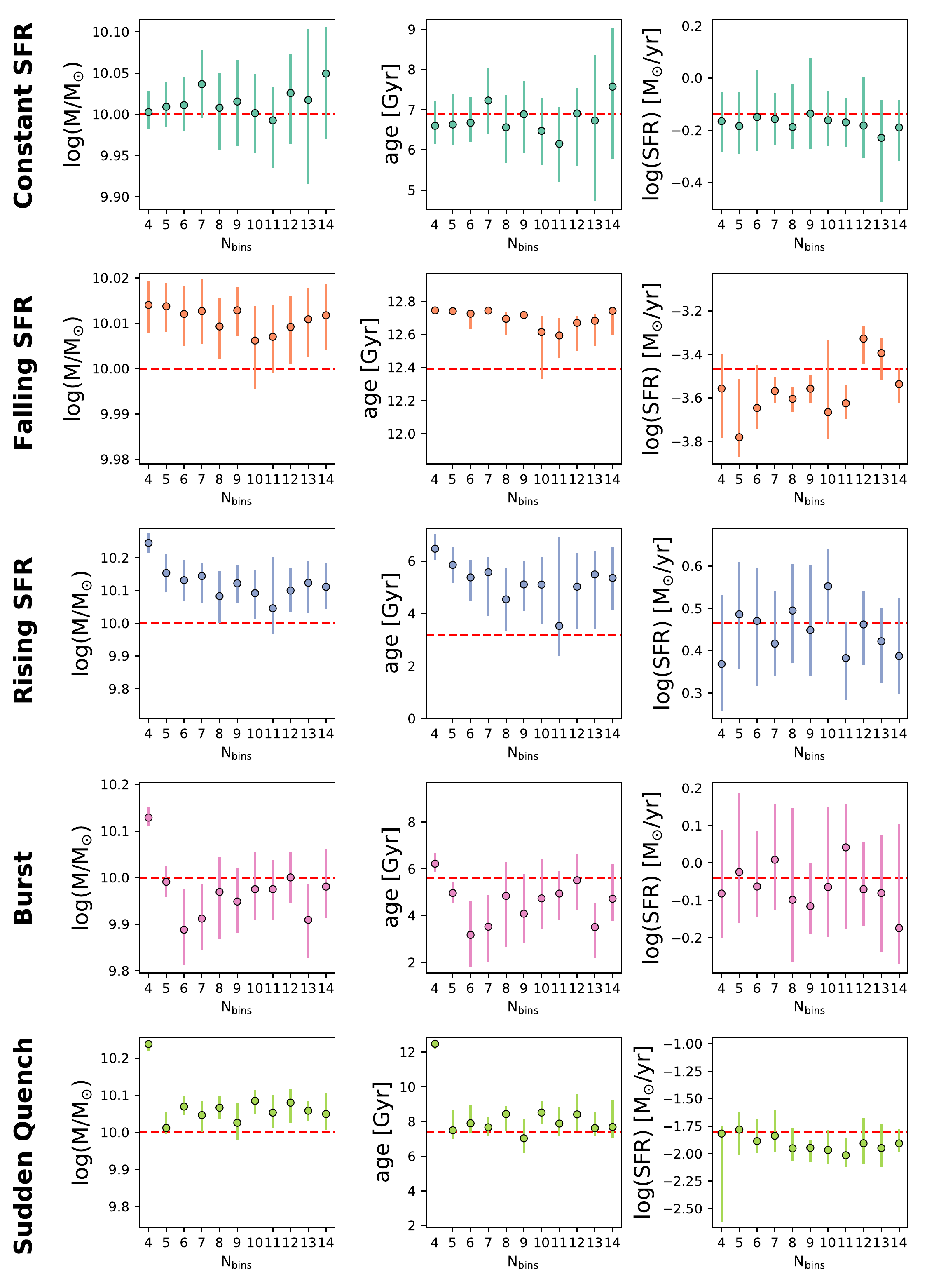}
\caption{Posterior masses, SFRs, and mass-weighted ages from fitting the mock photometry for a variety of input SFHs and number of SFH time bins. Each column shows the results from a different SFH prior, while each row shows a different parameter. The input values are shown with a dashed line. The SFR is calculated by averaging over the most recent 100 Myr.}
\label{fig:mock_timebin_dpars}
\end{center}
\end{figure*}

Figure \ref{fig:mock_timebin_sfh} shows the recovered SFHs as a function of N$_{\mathrm{bins}}$ while Figure \ref{fig:mock_timebin_dpars} shows the change in derived stellar mass, star formation rate, and mass-weighted age as a function of N$_{\mathrm{bins}}$. For N$_{\mathrm{bins}} \gtrsim 5$, there are no clear trends with N$_{\mathrm{bins}}$ in the star formation history posteriors or parameter posteriors, and the posteriors are largely consistent within their 1$\sigma$ error bars. This generally holds true across all of the input SFHs, with the exception of a few measurements for the SFR and age in the falling SFH. This is not systematic with N$_{\mathrm{bins}}$ and likely caused by inefficiencies in the sampler when sampling near the edge of the priors.

This confirms that the results presented here are not strongly affected by bin edge effects, or by overfitting. It also shows that using too few bins (N$_{\mathrm{bins}} \lesssim$ 4) fails to recover vital information in the SED. Notably, the effective prior on SFR(t) does depend on N$_{\mathrm{bins}}$: allowing more bins with a fixed prior on $\log({\mathrm{SFR}_{t_0} / \mathrm{SFR}_{t_0-\Delta t}})$ for each bin means a more flexible prior on SFR(t). However, this variation is too small to drive strong trends in the posteriors.
\onecolumngrid
\clearpage
\onecolumngrid
\clearpage
\twocolumngrid

\section{Tuning the continuity prior to star formation histories from Illustris}
\label{sec:illustris}
Here we use the ensemble of star formation histories from the Illustris hydrodynamical simulation to tune the prior on $\log({\mathrm{SFR}_{t_0} / \mathrm{SFR}_{t_0-\Delta t}})$ for the continuity model presented in Section \ref{sec:continuity}. These SFHs have been adopted from the data presented in \citet{diemer17} and includes $N=29203$ galaxies in the stellar mass range $9 < \log(\mathrm{M/M}_{\odot}) < 12.5$. The SFHs have 100 time steps spaced evenly in linear space and are based on the star formation model described in \citet{springel03}.

The SFHs cover all stars formed in a galaxy at $z = 0$, including all progenitors. For low-mass galaxies, the vast majority of stars is formed in-situ. For high-mass galaxies, however, a large fraction of their stars have formed in other halos/galaxies and merged with a more massive galaxies \citep{rodriguez16}. This means that the SFHs of high-mass galaxies at $z>0$ are actually reported as the sum of the SFHs of all of the galaxies which will eventually combine to form a single massive $z=0$ galaxy. This will bias the reported distribution to be somewhat narrower than the true distribution.

\begin{figure*}[t!h!]
\begin{center}
\includegraphics[width=0.85\linewidth]{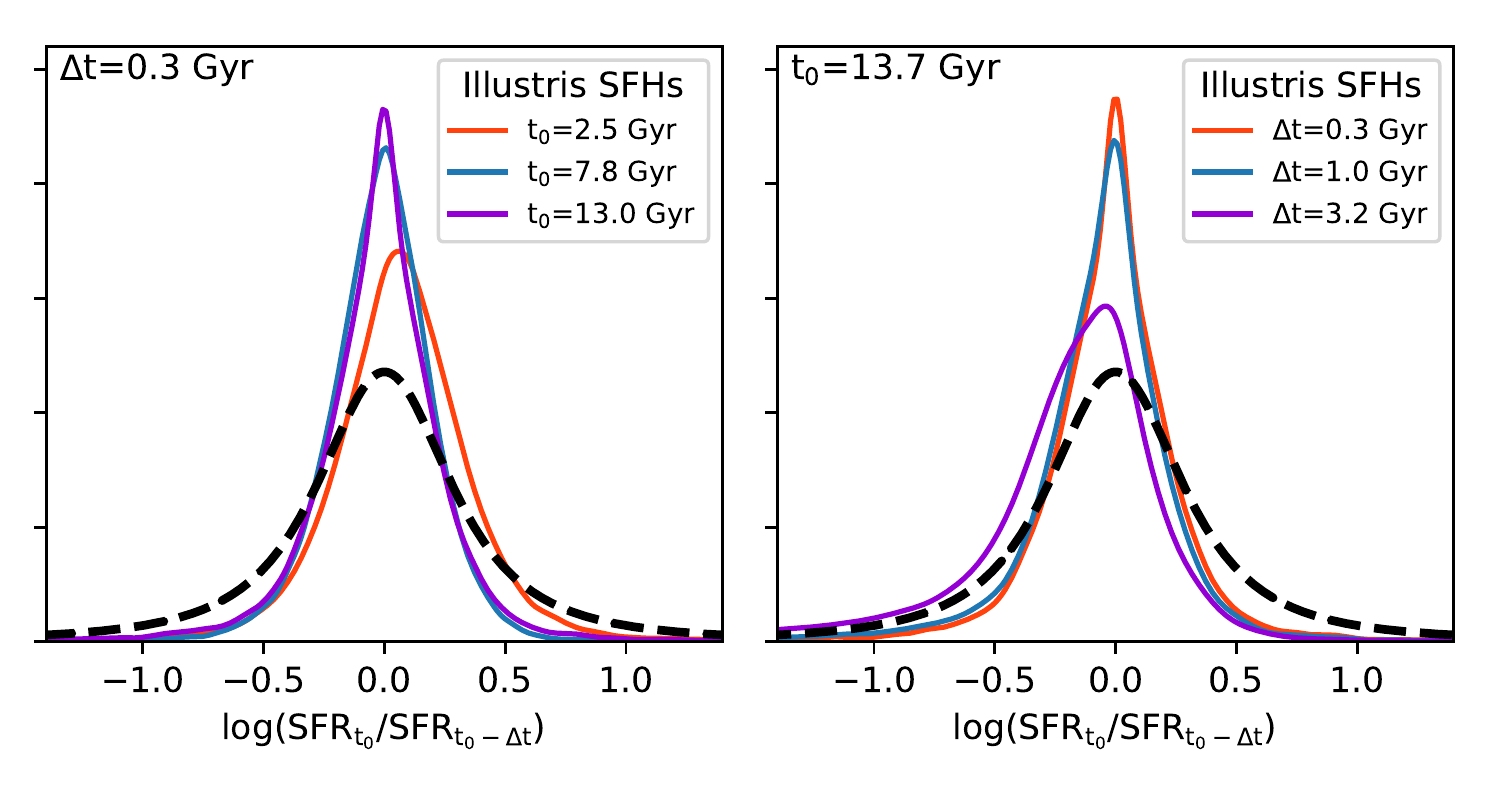}
\caption{The prior adopted for the continuity model (dashed black line) compared to $\log({\mathrm{SFR}_{t_0} / \mathrm{SFR}_{t_0-\Delta t}})$ measured from Illustris galaxies (colored lines). The left panel shows this distribution for several values of $t_0$ at fixed $\Delta t$ while the right panel has a variable $\Delta t$ and a fixed $t_0$. The adopted continuity prior is a reasonable replication of, though slightly broader than the distribution of Illustris SFHs for a variety of choices for $\Delta t$ and $t_0$.}
\label{fig:illustris}
\end{center}
\end{figure*}

Figure \ref{fig:illustris} compares the adopted Student's-t prior to the distribution of ratios of SFR$_{t_0}$/SFR$_{t_0-\Delta t}$ from the Illustris simulation for several values of $\Delta t$ and $t_0$. The adopted prior is a reasonable representation of the Illustris star formation histories at a variety of redshifts and differences in size for the fixed time bins. One of the striking differences between the prior and the simulation results is that the center of the distribution changes with $\delta t$, which reflects the simple finding that SFRs were higher at earlier times. We avoid including this effect in the prior so as to keep the findings general. However in principle, the cosmic rise and fall of galaxy star formation rates could be hard-coded into the prior to increase the accuracy of recovered galaxy SFHs.

\section{SFH recovery as a function of signal-to-noise}
\label{sec:sn_append}
\begin{figure*}[t!h!]
\begin{center}
\includegraphics[width=0.5\linewidth]{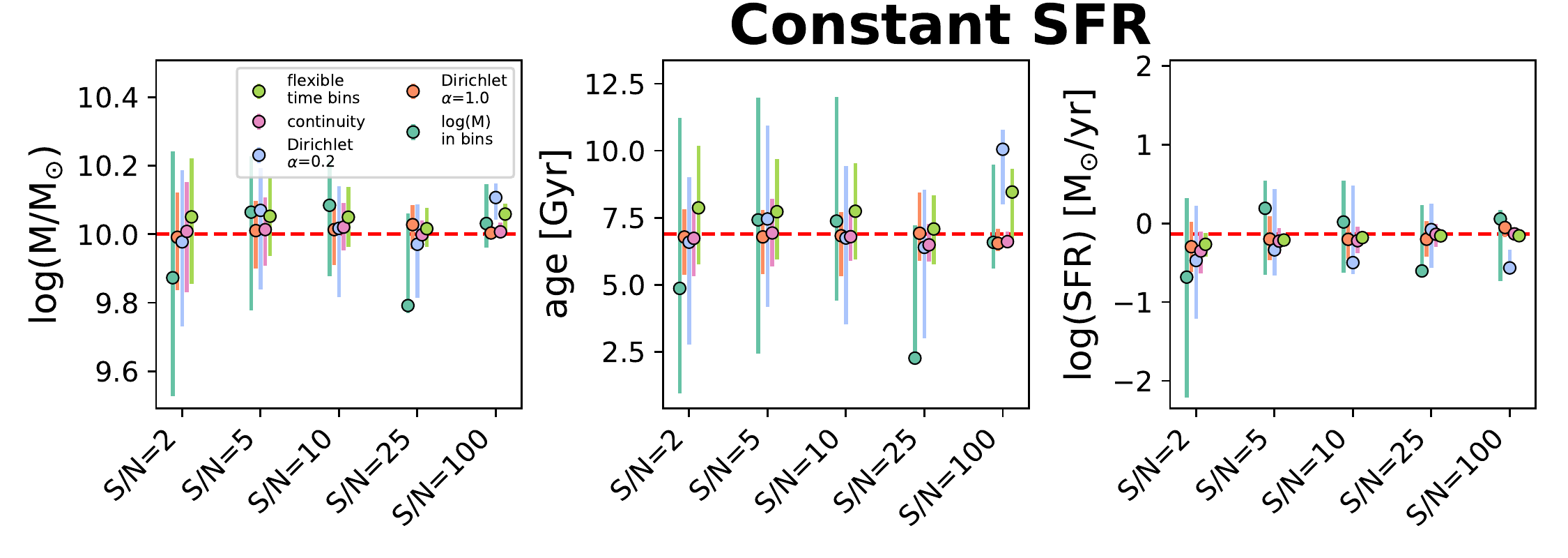}
\includegraphics[width=0.95\linewidth]{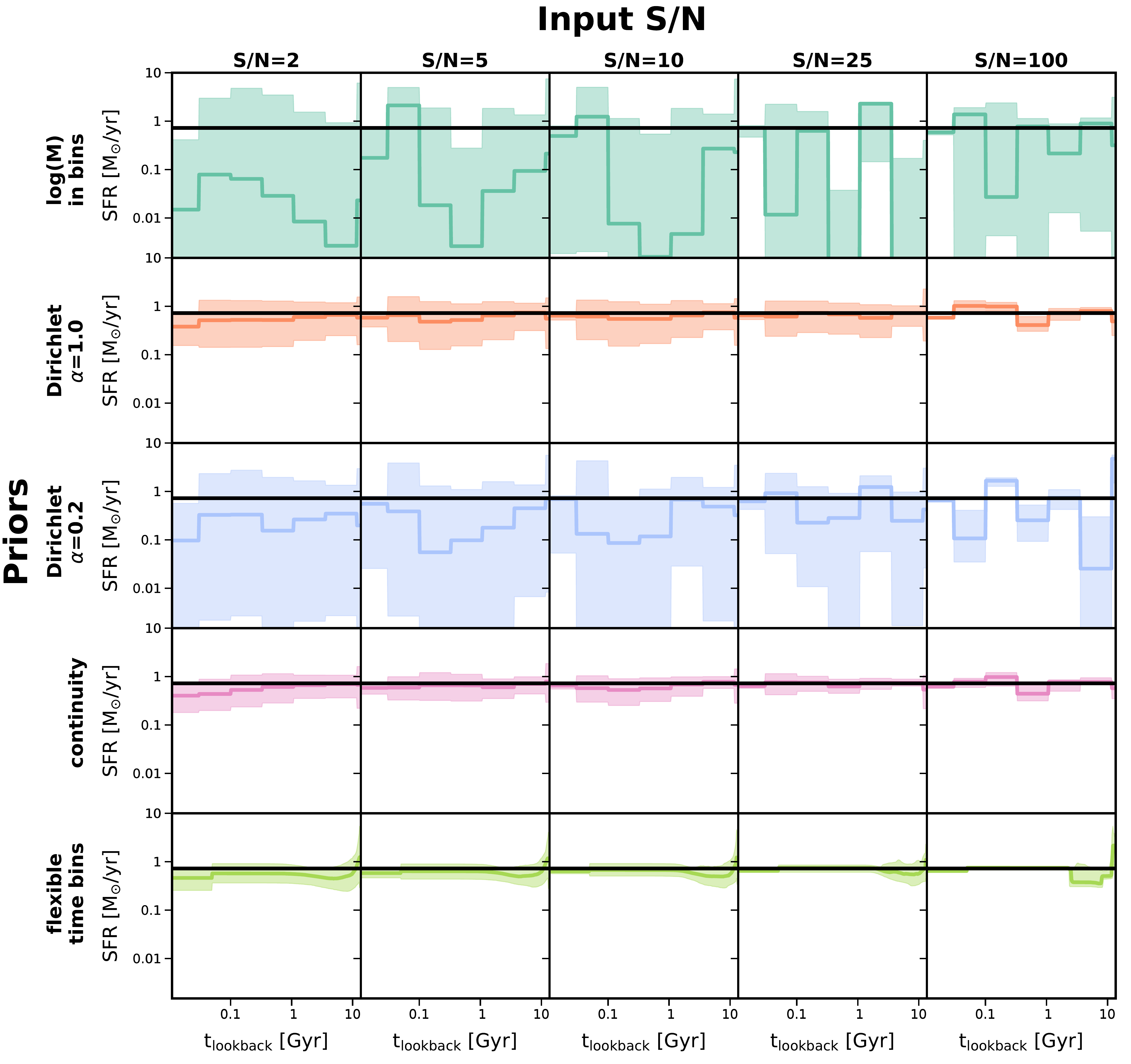}
\caption{The effect of signal-to-noise on the posteriors for different assumed priors for a constant input SFH. The top panels show the derived mass and age as a function of prior and S/N. The lower grid of panels shows the posterior SFHs. The input is a solid black line and the dashed black line is re-binned to match the nonparametric time bins. The median of the posterior is a thick colored line while the 16$^{\mathrm{th}}$-84$^{\mathrm{th}}$ percentiles are shown as shaded regions. Even with very precise S/N=100 photometry, the derived parameters and their associated error bars are sensitive to the chosen prior.}
\label{fig:sn_constant}
\end{center}
\end{figure*}

\begin{figure*}[t!h!]
\begin{center}
\includegraphics[width=0.5\linewidth]{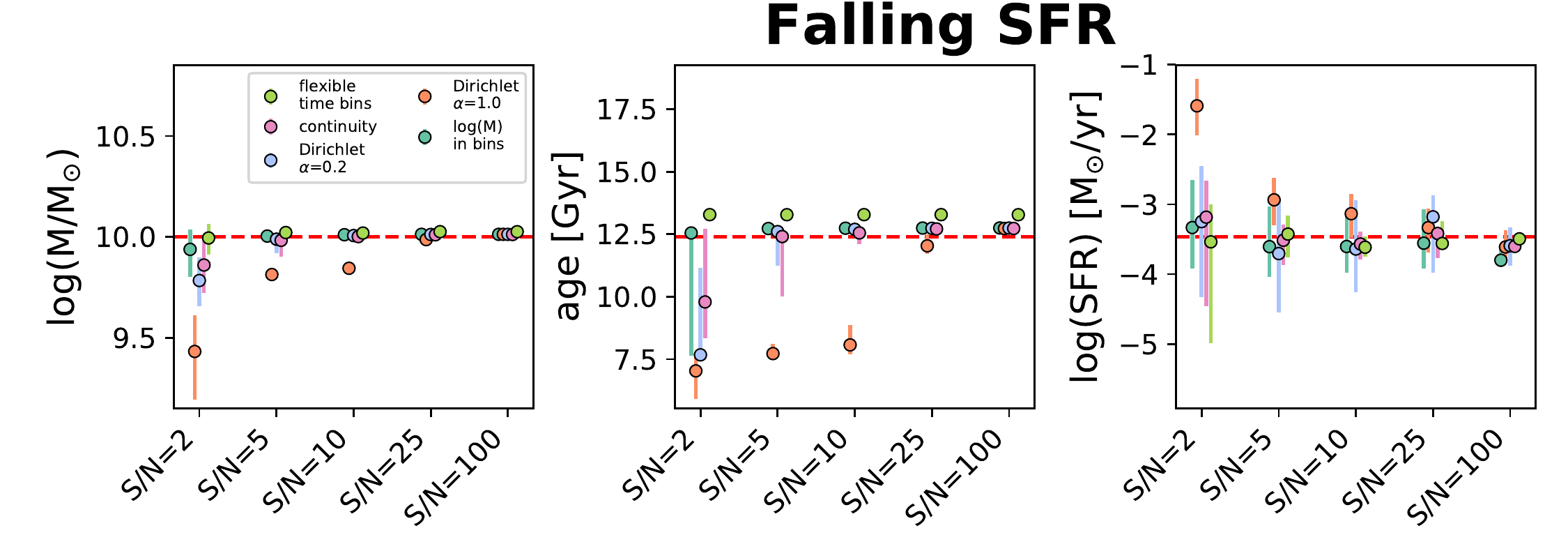}
\includegraphics[width=0.95\linewidth]{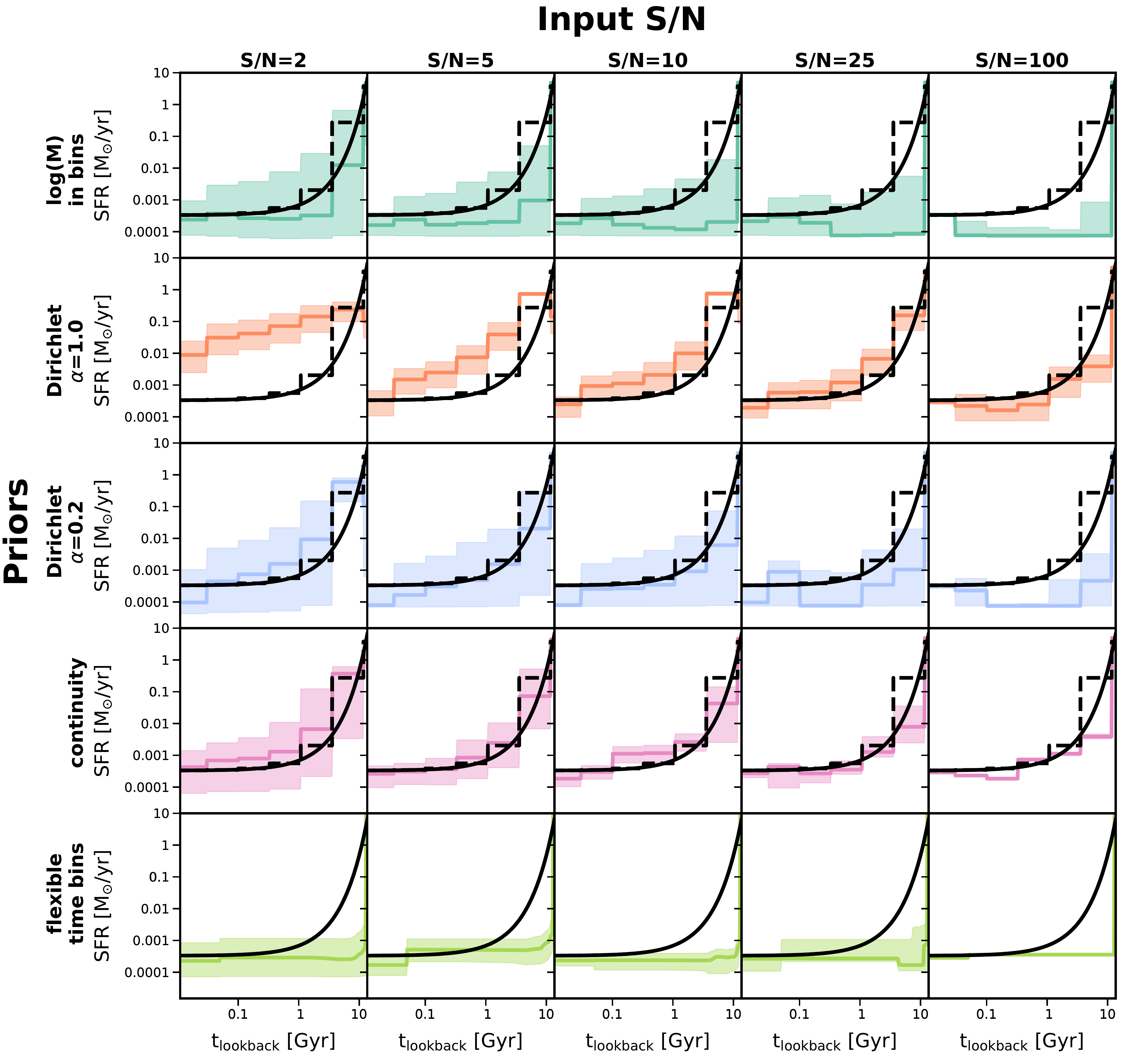}
\caption{The effect of signal-to-noise on the posteriors for different assumed priors for a falling input SFH. The top panels show the derived mass and age as a function of prior and S/N. The lower grid of panels shows the posterior SFHs. The input is a solid black line and the dashed black line is re-binned to match the nonparametric time bins. The median of the posterior is a thick colored line while the 16$^{\mathrm{th}}$-84$^{\mathrm{th}}$ percentiles are shown as shaded regions. Even with very precise S/N=100 photometry, the derived parameters and their associated error bars are sensitive to the chosen prior.}
\label{fig:sn_falling}
\end{center}
\end{figure*}

\begin{figure*}[t!h!]
\begin{center}
\includegraphics[width=0.5\linewidth]{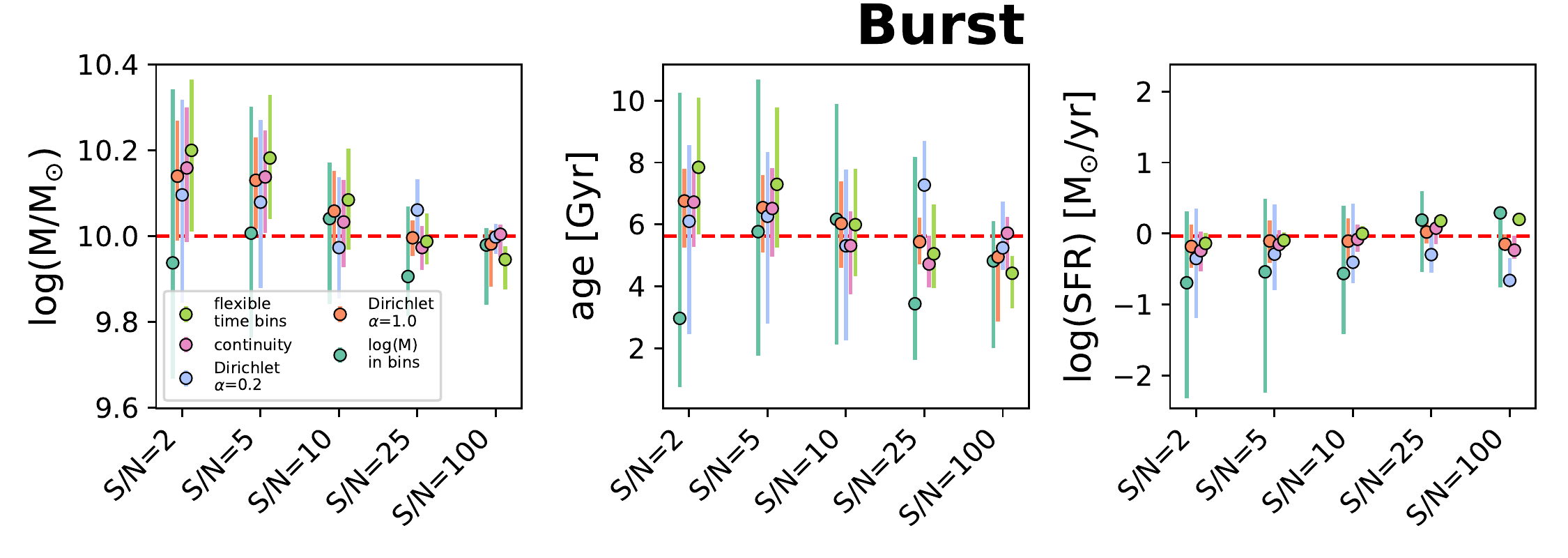}
\includegraphics[width=0.95\linewidth]{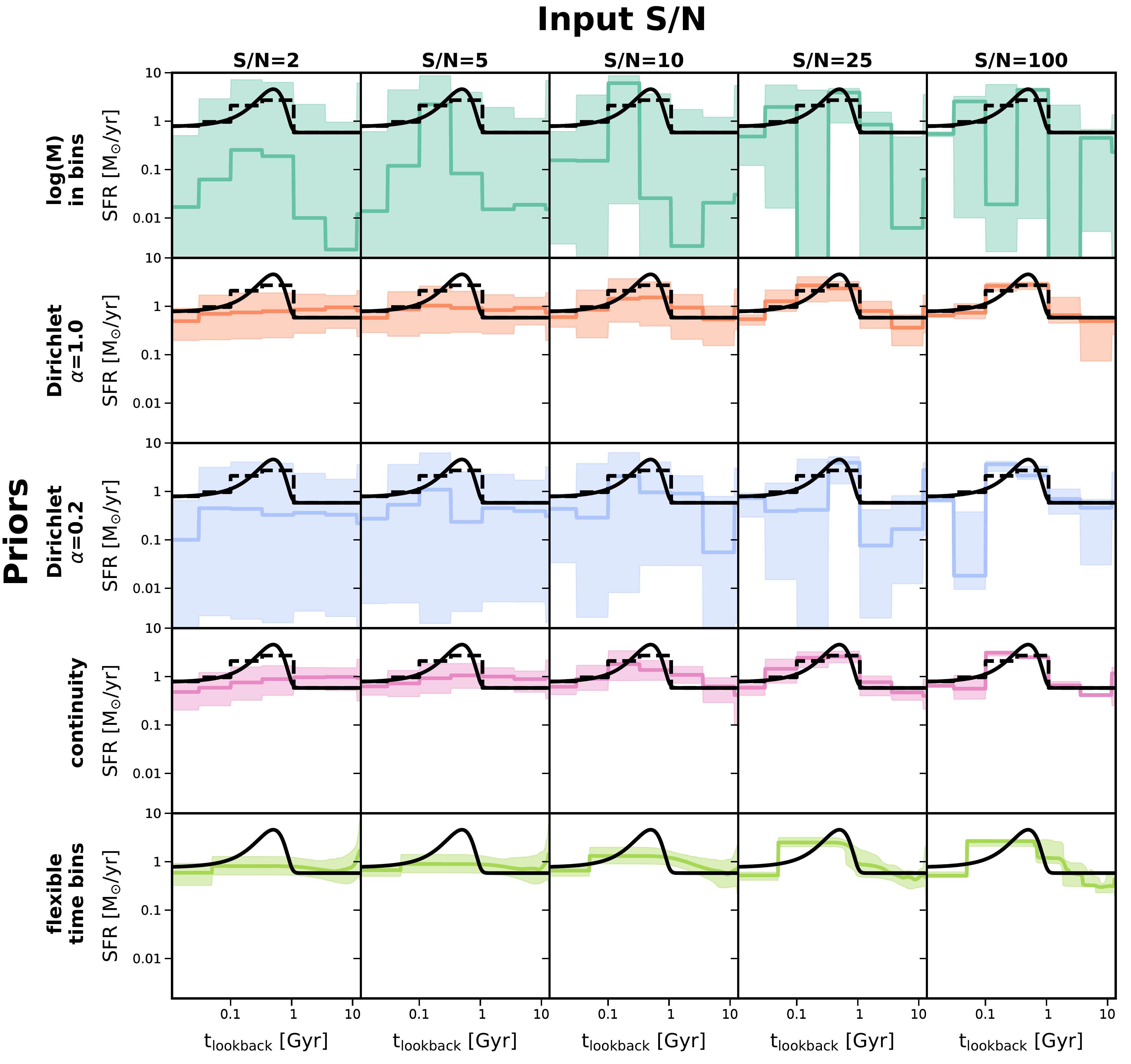}
\caption{The effect of signal-to-noise on the posteriors for different assumed priors for a constant+burst input SFH. The top panels show the derived mass and age as a function of prior and S/N. The lower grid of panels shows the posterior SFHs. The input is a solid black line and the dashed black line is re-binned to match the nonparametric time bins. The median of the posterior is a thick colored line while the 16$^{\mathrm{th}}$-84$^{\mathrm{th}}$ percentiles are shown as shaded regions. Even with very precise S/N=100 photometry, the derived parameters and their associated error bars are sensitive to the chosen prior.}
\label{fig:sn_burst}
\end{center}
\end{figure*}

\begin{figure*}[t!h!]
\begin{center}
\includegraphics[width=0.5\linewidth]{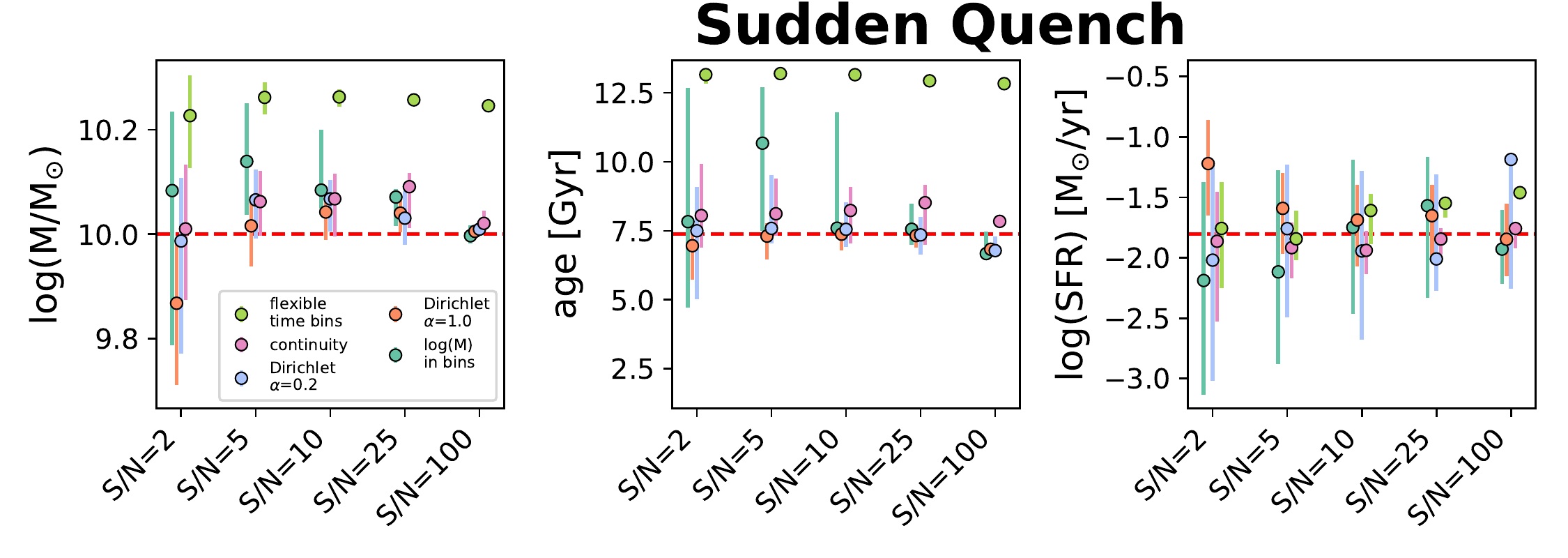}
\includegraphics[width=0.95\linewidth]{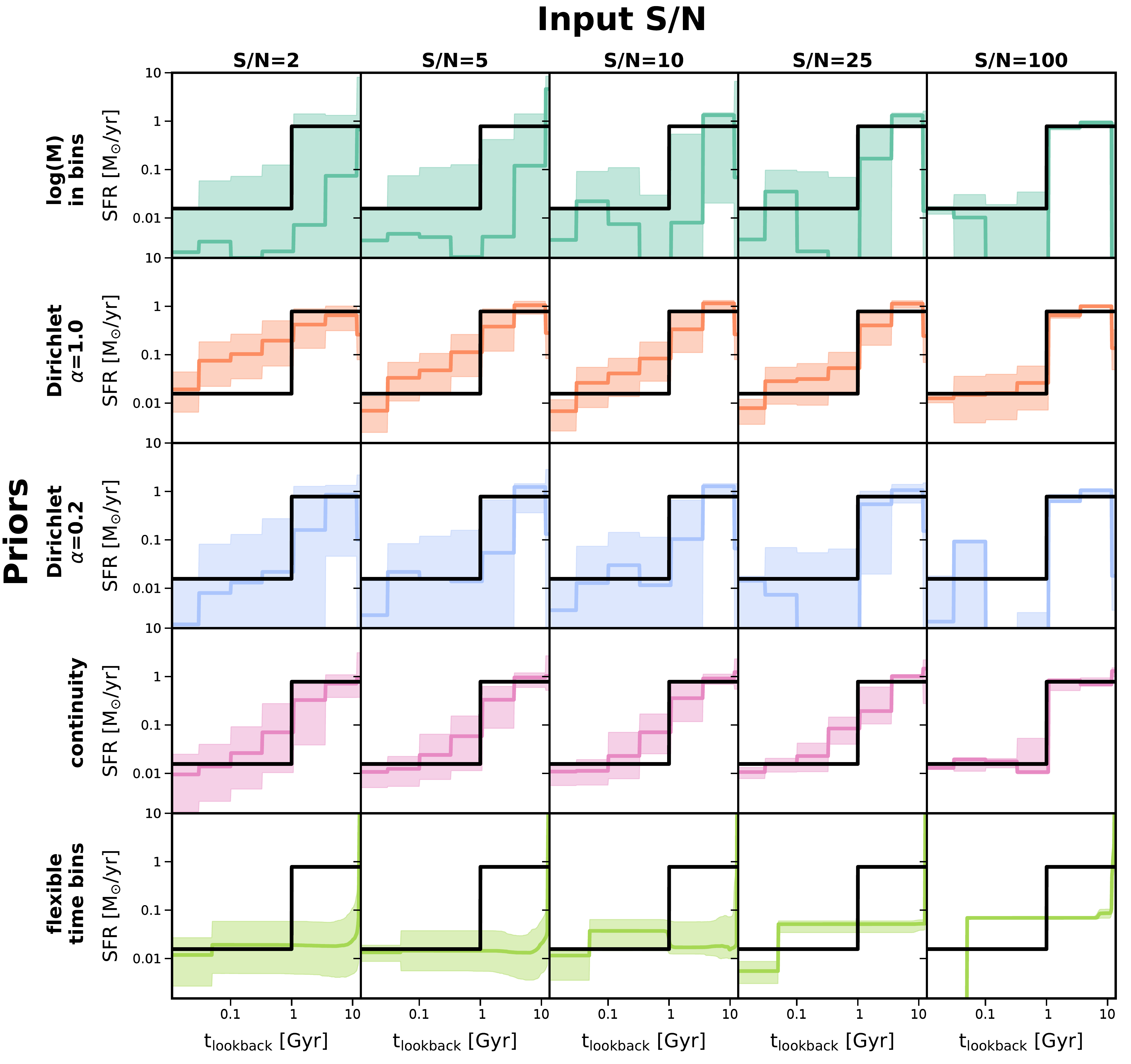}
\caption{The effect of signal-to-noise on the posteriors for different assumed priors for a constant input SFH with a sudden quenching event. The top panels show the derived mass and age as a function of prior and S/N. The lower grid of panels shows the posterior SFHs. The input is a solid black line and the dashed black line is re-binned to match the nonparametric time bins. The median of the posterior is a thick colored line while the 16$^{\mathrm{th}}$-84$^{\mathrm{th}}$ percentiles are shown as shaded regions. Even with very precise S/N=100 photometry, the derived parameters and their associated error bars are sensitive to the chosen prior.}
\label{fig:sn_quench}
\end{center}
\end{figure*}

In Figures \ref{fig:sn_constant}, \ref{fig:sn_falling}, \ref{fig:sn_burst}, and \ref{fig:sn_quench} we show the SFH recovery as a function of S/N for every mock input SFH. These figures are analogous to Figures \ref{fig:sn_trends_pars} and \ref{fig:sn_trends_sfh}.

\bibliography{/Users/joel/my_papers/tex_files/jrlbib}
\end{document}